\documentclass[reprint,prl,letterpaper,superscriptaddress]{revtex4-1}

\pdfoutput=1

\usepackage{amsmath,amssymb,amsthm,mathrsfs, cancel}
\usepackage{siunitx,mhchem}

\usepackage{graphicx}
\usepackage{wrapfig, subfig}

\usepackage{blindtext}
\usepackage{array,dcolumn}

\usepackage[all]{nowidow}

\usepackage{hyperref}
\hypersetup{
    colorlinks=true,
    linkcolor=blue,
    citecolor=blue,
    urlcolor=blue,
}

\newcommand \half {\frac{1}{2}}

\begin{document}

\title{Incorporation of random alloy \ce{GaBi_xAs_{1-x}} barriers in InAs quantum dot molecules~(I): energy levels and confined hole states.}

\author{Arthur Lin}
\email{arthur.lin@nist.gov}
\affiliation{Joint Quantum Institute, \\
 University of Maryland and National Institute of Standards and Technology,
 College Park, Maryland 20742, USA
}

\author{Matthew F.\ Doty}
\affiliation{Department of Materials Science and Engineering, 
 University of Delaware,
 Newark, Delaware 19716, USA
}

\author{Garnett W.\ Bryant}
\affiliation{Joint Quantum Institute, \\
 University of Maryland and National Institute of Standards and Technology,
 College Park, Maryland 20742, USA
}
\affiliation{Nanoscale Device Characterization Division, \\
 National Institute of Standards and Technology,
 Gaithersburg, Maryland 20899, USA
}

\date{\today}


\begin{abstract}
Self-assembled InAs quantum dots (QDs), which have long hole-spin coherence times and are amenable to optical control schemes, have long been explored as building blocks for qubit architectures. One such design consists of vertically stacking two QDs to create a quantum dot molecule (QDM) and using the spin-mixing properties of ``molecule-like'' coupled hole states for all-optical qubit manipulation. In this paper, the first of two papers, we introduce the incorporation of dilute GaBi$_x$As$_{1-x}$ alloys in the barrier region between the two dots. GaBi$_x$As$_{1-x}$ is expected to increase the spin-mixing of the molecular states needed for qubit operations by raising the barrier valence band edge and spin-orbit splitting. Using an atomistic tight-binding model, we compute the properties of GaBi$_x$As$_{1-x}$ and the modification of hole states that arise when the alloy is used in the barrier of an InAs QDM. An atomistic treatment is necessary to correctly capture non-traditional alloy effects such as the band-anticrossing valence band. It also allows for the study of configurational variances and clustering effects of the alloy. We find that in InAs QDMs with a GaBiAs interdot barrier, electron states are not strongly affected by the inclusion of Bi. However, hole states are much more sensitive to the presence and configuration of Bi in the barriers. By independently studying the alloy-induced strain and electronic scattering off Bi and As orbitals, we conclude that an initial increase in QDM hole state energy at low Bi concentration is caused by the alloy-induced strain. We further find that the decrease in QDM hole energy at higher Bi concentrations can only be explained when both alloy strain and orbital effects are considered. In our second paper, we use the understanding developed here to discuss how the alloyed barriers contribute to enhancement in hole spin-mixing and the implications for QDM qubit architectures.
\end{abstract}

\maketitle


\section{Introduction} \label{sec:intro}

\paragraph*{}
Semiconductor quantum computing has gained much attention in recent years.  A variety of systems
have demonstrated the capability to effectively isolate and control individual spins and their interactions \cite{ladd_quantum_2010,morton_embracing_2011,awschalom_quantum_2013}, a crucial step toward designing an architecture for quantum information processing.  Hole spins in self-assembled semiconductor quantum dots (QDs) offer intriguing possibilities as a potential qubit architecture.  Most importantly, self-assembled QDs generally have deeper potential wells than other semiconductor architectures such as electrostatically defined quantum dots and silicon dopants, allowing QDs to operate at higher temperatures \cite{ladd_quantum_2010}, even room temperature.  Second, the optical nature of these systems allows for ultrafast control not available in gated systems \cite{bonadeo_coherent_1998}.  While the choice of III-V semiconductors over Si means that there are interactions with nuclear spins that shorten coherence times \cite{merkulov_electron_2002,khaetskii_spin_2000,khaetskii_electron_2002}, increased coherence time can be achieved by using holes spins as qubits \cite{brunner_coherent_2009}.  The $p$-like orbital of holes, unlike the $s$-like orbital of electrons, has little probability to be at the atomic nuclei \cite{eble_holenuclear_2009,testelin_hole-spin_2009}.  The reduced probability results in decreased interaction with nuclear spin, increasing the hyperfine coherence time for holes to \SI{10}{\nano\second}, an order of magnitude longer than that of electrons \cite{testelin_hole-spin_2009}.

\paragraph*{}
One potential shortcoming of self-assembled dots is the inevitable structural inhomogeneity of the array of QDs used in a qubit architecture.  Tuning many QDs to a fixed laser frequency or photonic cavity mode is a practical challenge. However, as recently proposed by Economou \textit{et al.} \cite{economou_scalable_2012}, this issue can be overcome by the use of vertically stacked quantum dot molecules (QDM), in which pairs of InAs QDs separated by a GaAs tunnel barrier are stacked along the growth axis, nearly aligned.  When the QDs are closely spaced, electric fields from a diode structure can be used to tune the hole states from the pair of QDs in and out of resonance, controlling the coupling strength between the dots and forming delocalized molecular states \cite{bayer_coupling_2001,krenner_direct_2005,ortner_control_2005,stinaff_optical_2006,doty_optical_2008,doty_antibonding_2009,doty_hole-spin_2010,liu_situ_2011}.  In this approach, the hole spin in the larger dot (typically the top dot), provides the spin-up and spin-down hole spin states that serve as the binary qubit basis.  The QDM is tuned well away from resonance so that these two states have well-defined spin and are strongly localized to the top dot.  Indirect optical transitions, ones that drive the hole states ($ h^+ $) to specially engineered charged trion states ($ X^+ $), are used to provide all-optical initialization, manipulation and readout of the hole spin qubit. The particular $ X^+ $ states are driven by indirect transitions because the electron is in the top dot but the second hole is in a molecular state spread between the two dots.  The electric field provided by the first hole serves to put the optically excited hole into a molecular state in resonance between the two dots.  Indirect optical transitions have much larger Stark shifts than direct transitions and thus have a much larger tuning range.

\paragraph*{}
Economou's schemes for initialization, manipulation, and readout are shown schematically in Fig.~\ref{fig:qubit_control} \cite{economou_scalable_2012}.  Readout involves indirect resonant transitions between the hole spin and a trion state that preserve spin.  The readout is nondestructive because no transverse magnetic field, which would lead to spin-mixing, is used in any of the qubit operations.  Spin initialization and manipulation both depend on the molecular hole state being spin mixed by the tunnel coupling.  The existence and origin of this hole spin-mixing for molecular hole states has been established and explored \cite{doty_hole-spin_2010,rajadell_large_2013,planelles_symmetry-induced_2015}, as will be discussed in more detail below.  Spin initialization occurs via optical shelving, with a resonant, indirect, optical transition taking one hole spin to a $ X^+ $ state which, because of the spin-mixing of the hole state, can then decay back to both spin states of the hole.  The spin becomes initialized to the lower energy state after multiple transitions.  Coherent spin manipulation is driven by two nonresonant, indirect optical Raman transitions which couple the two spin states of the qubit to the spin-mixed trion states.  Increased spin mixing leads to faster spin initialization and more effective coherent control.

\begin{widetext}
\onecolumngrid
\begin{figure}[ht]
  \centering
  \includegraphics[width=0.9\textwidth]{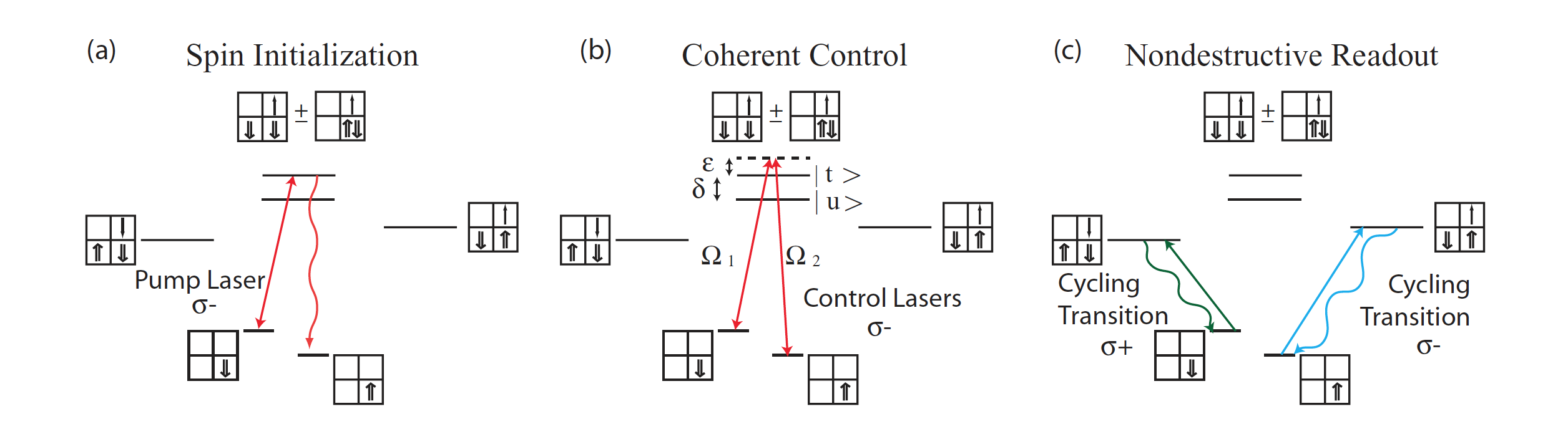}
    \caption[Proposed strategy for qubit control using optical transitions of QDMs.]{\small Proposed strategy for (a) spin initialization, (b) manipulation, and (c) readout using optical transitions of QDMs.  Top/bottom rows indicate the spin projections of electrons/holes in the top (right column) and bottom (left column) QDs.  (Excerpt from Economou \textit{et al.}, Ref.\ \cite{economou_scalable_2012})}
  \label{fig:qubit_control}
\end{figure}
\end{widetext}

\paragraph*{}
The magnitude of the spin-mixing is determined by the alignment of the two dots.  When the two dots are perfectly aligned, there is no spin mixing.  However, the dots are typically offset in structures grown by molecular beam epitaxy.  When this is the case, rotational symmetry is broken (i.e., a lateral lever arm exists), the states acquire angular momentum, and the induced spin-orbit coupling leads to spin mixing \cite{doty_electrically_2006,doty_optical_2008,climente_theory_2008,doty_hole-spin_2010,rajadell_large_2013,planelles_symmetry-induced_2015,ma_hole_2016}.  Larger lateral offsets lead to larger spin-mixing.  Unfortunately, the lateral offsets typically found in grown QDMs only provide small spin-mixing \cite{climente_theory_2008,doty_hole-spin_2010,rajadell_large_2013,ma_hole_2016}.  A lateral electric field that is asymmetric between the two dots in the QDM can mimic the effects of a lateral offset to enhance spin-mixing \cite{ma_hole_2016}.  However, such an effort could be difficult to engineer within photonic structures.  In this paper and a follow-up paper, we study how enhanced spin-mixing can be engineered by changing the barrier composition from GaAs to GaBiAs.  In this paper, we develop and employ an atomistic tight-binding (TB) formalism to compute the energy levels of electrons and holes confined in QDM structures including GaBiAs barriers.  In the follow-up paper, we apply these results to understand the resulting enhancement in hole spin-mixing and the implications for QDM qubit architectures.

\begin{figure}[ht]
  \centering
  \includegraphics[width=0.4\textwidth]{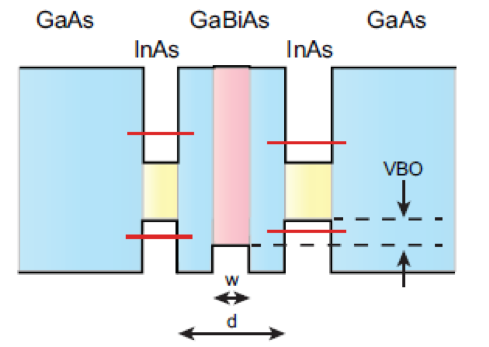}
    \caption[Conceptual energy structure of \ce{InAs} QDM, with proposed \ce{GaBiAs} in barrier region.]{\small (Color online) Conceptual schematic of the band edges of an \ce{InAs} QDM, with \ce{GaBiAs} in the barrier region. (VBO: valence band offset.)}
  \label{fig:system_schematic}
\end{figure}

\paragraph*{}
Recently, random alloy GaBiAs has generated substantial interest \cite{janotti_theoretical_2002,yoshida_temperature_2003,zhang_similar_2005,alberi_valence-band_2007,deng_band_2010,usman_tight-binding_2011,batool_electronic_2012,usman_impact_2013,bannow_configuration_2016,bannow_valence_2017}.  The addition of dilute amounts of Bi to GaAs allows the valence band to be tuned, through band anticrossing (BAC) effects, to raise the valence-band edge (VBE) while minimally affecting the conduction band (CB) \cite{zhang_similar_2005,usman_tight-binding_2011}.  Incorporating \ce{GaBi_{x}As_{1-x}} into the region separating the two dots of a QDM (as shown in Fig.~\ref{fig:system_schematic}) lowers the interdot barrier for holes and enhances hole tunneling, while preserving the electron band structure.  Additionally, literature suggests that GaBiAs has much stronger intrinsic spin-orbital (SO) interactions compared to GaAs because Bi is a heavier atom \cite{zhang_similar_2005,usman_tight-binding_2011,batool_electronic_2012}, which could also enhance the hole spin-mixing in QDMs. 

\paragraph*{}
While GaAs is a typical III--V direct gap semiconductor, GaBi is a negative-gap semiconductor (semimetal).  Moreover, Bi has substantially larger spin-orbit coupling than As.  GaBiAs alloys are expected to be significantly different from typical III--V alloys and to have significant local spin-orbit interaction that may provide new tools for modifying the hole spin physics of InAs QDMs.  The purpose of this paper is to investigate how GaBiAs alloys should be treated to correctly model their properties in InAs QDMs and to determine which effects of the alloy barriers influence the QDM states.  We assess the sensitivity of the QDM states to random fluctuations in the alloy configuration as well as Bi clustering in the alloy barrier.  Our results show that an atomistic treatment is necessary to properly account for the alloy effects.  We separately investigate the effects of (1) the strain induced when the larger Bi replaces the smaller As and (2) the effects when the As energy levels are replaced by Bi orbitals and energy levels, i.e., the scattering induced when Bi replaces As. We find that the induced strain is a global effect, while the difference between Bi and As orbitals is more local.  In a second paper, we report how hole spin-mixing and tunnel coupling are affected when GaBiAs interdot barriers are used and how the alloy barriers can be utilized to enhance the operation of hole spin qubits in InAs QDMs.


\section{Computational Model} \label{subsec:gabias-methods}

\paragraph*{}
We model a QDM consisting of two vertically stacked InAs QDs.  Experimentally, the QDM heterostructure would be formed by molecular-beam epitaxial deposition of two consecutive layers of self-assembled InAs QDs, separated by a thin layer of GaAs/GaBiAs acting as a barrier.  When the second layer of InAs is deposited above the GaAs/GaBiAs barrier, the strain induced by the lattice mismatch between the dot in the first InAs layer and the barrier GaAs/GaBiAs provides a nucleation point for a second dot to form directly above the first dot \cite{wasilewski_size_1999}. The two InAs dots are thus naturally stacked vertically (along the growth direction); the substrate under the first dot and the cap above the second dot are GaAs; the barrier is engineered to be various GaBiAs/GaAs structures.  Details on the growth of these structures can be found in references \cite{scheibner_essential_2009,wasilewski_size_1999}.

\paragraph*{}
We model these QDMs using an $ sp^3s^* $ nearest-neighbor TB Hamiltonian \cite{harrison_bond-orbital_1973,harrison_bond-orbital_1974,pantelides_structure_1975,vogl_semi-empirical_1983,graf_electromagnetic_1995,ma_hole_2016}.  To account for strain from lattice mismatch, the Hamiltonian is constructed for a lattice whose atomic positions are relaxed via a valence force field (VFF) model.  The states and wave functions of interest are found by iterative diagonalization of the Hamiltonian, done with the ARPACK implementation of the Arnoldi method.  This allows us to find electron and hole states near the band edge.

\begin{table}[ht]
  \centering
  \caption[VFF parameters.]{\small Bond length and VFF parameters used for lattice relaxation. $ \alpha $ and $ \beta $ are in \si{\newton\per\meter}.  Bond lengths are in angstroms.} \label{table:vff_para}
  \begin{ruledtabular}
  \begin{tabular}{l|ddd}
    & \multicolumn{1}{r}{InAs\footnotemark[1]} & \multicolumn{1}{r}{GaAs\footnotemark[1]} & \multicolumn{1}{r}{GaBi\footnotemark[2]} \\
    \colrule
    $ \alpha $ (\si{\newton\per\meter})	&  35.18  &  41.49  & 34.0 \\
    $ \beta $ (\si{\newton\per\meter})	&  5.49  &  8.94  &  5.0 \\
    $ d_0 $ (\si{\angstrom})	& 2.622	& 2.448	& 2.740 \\
  \end{tabular}
  \end{ruledtabular}
  \footnotetext{Values from Zunger \textit{et al.\ }\cite{pryor_comparison_1998}}
  \footnotetext{Values converted from O'Reilly \textit{et al.\ }\cite{usman_tight-binding_2011}}
\end{table}

\paragraph*{}
The VFF parameters and bond lengths for the relevant materials are given in Table~\ref{table:vff_para}.  The parameters for InAs and GaAs are taken from Zunger \textit{et al.\ }\cite{pryor_comparison_1998}, and the parameters for GaBi are taken from O'Reilly \textit{et al.\ }\cite{usman_tight-binding_2011} (with lattice constant converted to bond length using a zincblende structure).  The TB parameters for InAs and GaAs, shown in Table~\ref{table:tb_para}, are taken from Vogl \textit{et al.\ }\cite{vogl_semi-empirical_1983}, with proper spin-orbit coupling parameters added. The TB parameters for GaBi are converted from O'Reilly's $ \sigma $ and $ \pi $-bond model \cite{usman_tight-binding_2011} to a $ p_x $ and $ p_y $ model.

\begin{table}[ht]
  \centering
  \caption[TB parameters.]{\small Tight-binding parameters (orbital energies, inter-atomic interaction, spin-orbit coupling and valence band offset) for an $ sps^* $ Hamiltonian.  A subscript $ x $ indicates a $ p $-orbital parallel to a crystal axis.  A subscript $ xx $ indicates two parallel $ p $-orbitals; a subscript $ xy $ indicates two perpendicular $ p $-orbitals.  The subscript $ a $ and $ c $ indicate anion and cation, respectively.  All units in \si{\electronvolt}.} \label{table:tb_para}
  \begin{ruledtabular}
  \begin{tabular}{l|ddd} 
    & \multicolumn{1}{r}{InAs\footnotemark[1]} & \multicolumn{1}{r}{GaAs\footnotemark[1]} & \multicolumn{1}{r}{GaBi\footnotemark[2]} \\
    \colrule
    $E_{s,a}$	& -9.5381	& -8.3431	& -8.3774 \\
    $E_{s,c}$	& -2.7219	& -2.6569	& -5.6126 \\
    $E_{x,a}$	&  0.9099	&  1.0414	&  0.1256 \\
    $E_{x,c}$	&  3.7201	&  3.6686	&  1.694 \\
    $E_{s^*,a}$	&  7.4099	&  8.5914	&  6.1262 \\
    $E_{s^*,a}$	&  6.7401	&  6.7386	&  5.8164 \\
    $V_{ss,ac}$		& -5.6052	& -6.4513	& -5.3700 \\
    $V_{sx,ac}$		&  3.0354	&  4.4800	&  5.4426 \\
    $V_{sx,ca}$		&  5.4389	&  5.7839	&  2.7771 \\
    $V_{xx,ac}$		&  1.8398	&  1.9546	&  0.9727 \\
    $V_{xy,ac}$		&  4.4693	&  5.0779	&  3.5143 \\
    $V_{s^*x,ac}$	&  3.3744	&  4.8422	&  4.1687 \\
    $V_{s^*x,ca}$	&  3.9097	&  4.8077	&  1.1778 \\
    \colrule
    $U_{so,a}$	&  0.140	&  0.140	&  0.672 \\
    $U_{so,c}$	&  0.000	&  0.000	&  0.03843 \\
    \colrule
    VBO	&	0.0	& -0.2 & 0.9 \\
  \end{tabular}
  \end{ruledtabular}
  \footnotetext{Values from Vogl \textit{et al.\ }\cite{vogl_semi-empirical_1983}}
  \footnotetext{Values converted from O'Reilly \textit{et al.\ }\cite{usman_tight-binding_2011}}
\end{table}

\paragraph*{}
To construct our lattice with a QDM and a GaBiAs interdot barrier, we start with a rectangular box of pure GaAs.  A pseudo-random number generator is used to select and replace As atoms with Bi in the interdot barrier.  Finally, the two QDs and their wetting layers are created by replacing the appropriate Ga atoms with In.  The hopping parameters between two atoms are given directly from Table~\ref{table:tb_para}.  Because an atom may be surrounded by different nearest neighbors, the orbital energies at the atomic site are taken to be the average of the orbital energies it has for each nearest neighbor, weighted by the number of each neighbor.  At an interface between InAs and GaBiAs, there may be occurrences of InBi.  
These will be few, so we use the parameters of InAs to describe InBi, as there is no known TB parameter set for InBi.  We do not expect the difference between InAs and InBi to alter our results in any significant manor.

\begin{figure}[ht]
  \centering
  \subfloat[Full GaBiAs Barrier]{
    \includegraphics[width=0.23\textwidth]{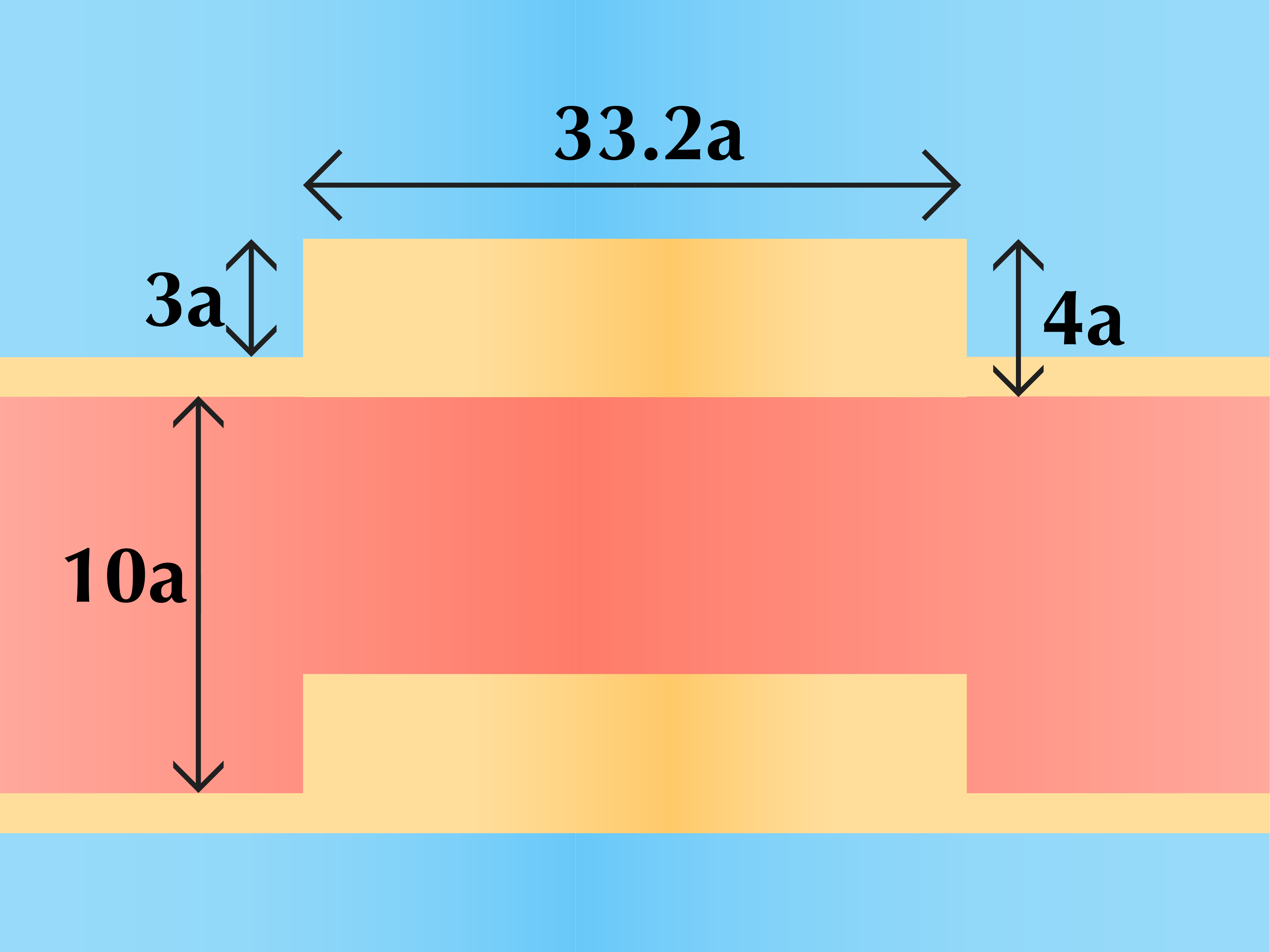}
    \label{fig-sub:system_geo-full}
  }
  \subfloat[Partial GaBiAs Barrier]{
    \includegraphics[width=0.23\textwidth]{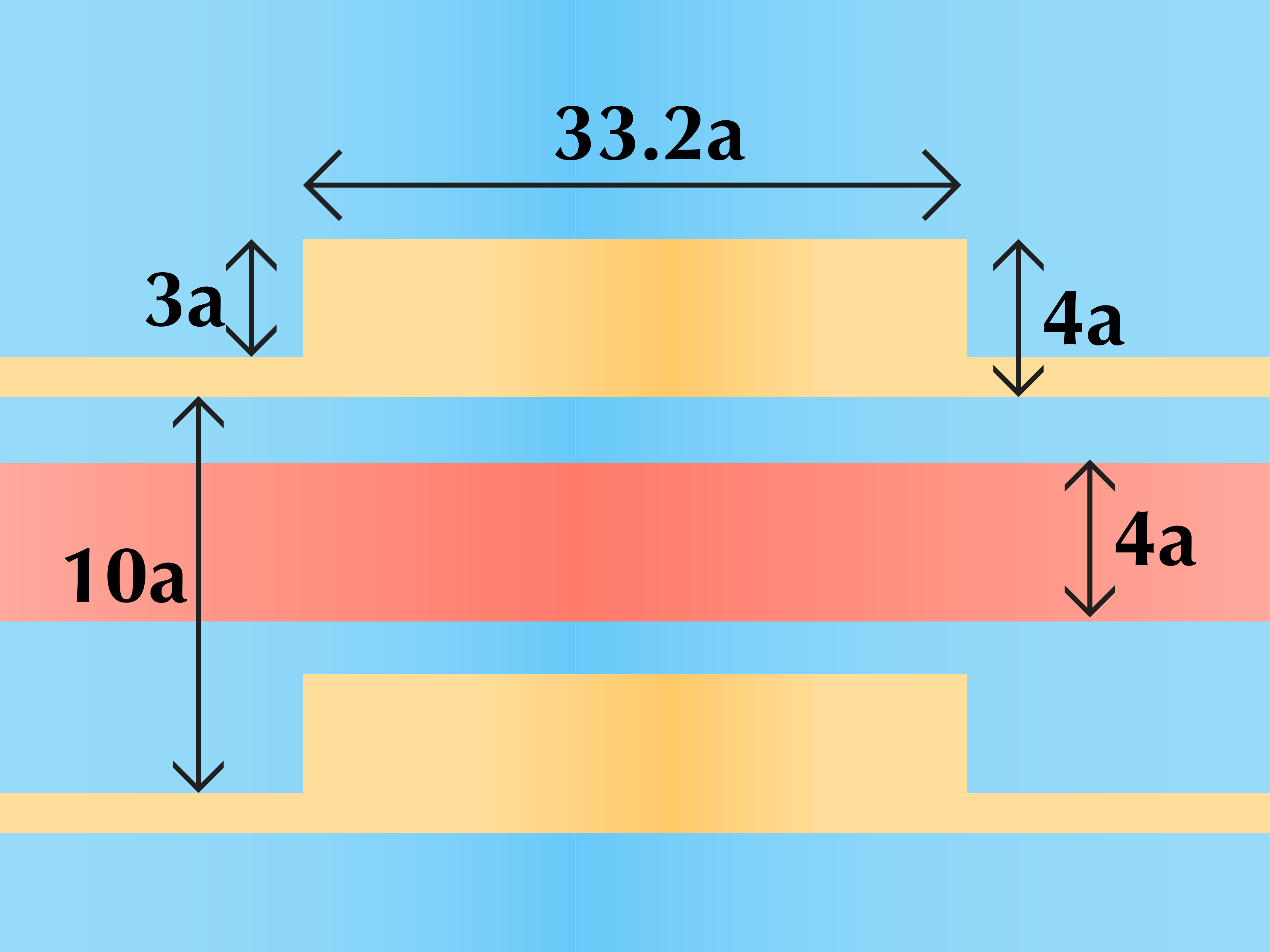}
    \label{fig-sub:system_geo-part}
  }
  \caption[General geometry of \ce{InAs} QDM, with proposed \ce{GaBiAs} in barrier region.]{\small (Color online) Schematic representation of system geometry.  GaAs in blue; InAs in yellow; \ce{GaBiAs} (a) spanning the full interdot region (b) as a partial layer inside the barrier, in pink.  ($x$ and $z$ axes not to scale.)}
  \label{fig:system_geo}
\end{figure}

\paragraph*{}
All calculations are carried out for cylindrical dots of radius $ 16.6a $, height $ 3a $ (not including the wetting layer), and $ 1a $ thick wetting layers ($a$ being the lattice constant), as shown in Fig.~\ref{fig:system_geo}.  The dots are perfectly aligned on top of one another and have an interdot distance (top of lower dot to the bottom of upper wetting layer) of $ 7a $.  A lattice of size $ 144a \times 144a \times 45a $ is used for calculating the strain relaxation via VFF.  After the new atomic positions are calculated, a $ 50a \times 50a \times 34a $ cutout of the relaxed lattice is made.  We construct our Hamiltonian from the smaller cutout to save significant computation time during diagonalization.  In construction of the TB Hamiltonian, hopping TB parameters are scaled by the relaxed bond length $ d $ as $ \left( \frac{d_0}{d} \right)^\eta $, where $ d_0 $ is the natural (unrelaxed) bond length of the respective material.  $ \eta $ for all materials is taken to be 2.9, typical for the class of materials of interest.  The particular value of 2.9 was chosen as it has historically agreed well with experimental results.  Spin-orbit parameters are also on-site parameters and therefore incorporated into the Hamiltonian as an average between the materials formed with the nearest-neighbors.

\section{Band Structure of G\lowercase{a}B\lowercase{i}A\lowercase{s}} \label{sec:intro-gabias_bbs}

\paragraph*{}
To understand QDMs with \ce{GaBi_{x}As_{1-x}} barriers, we must first understand the properties of \ce{GaBiAs}.  Detailed local-density approximation corrected (LDA+C), $ k \cdot p $, and semi-empirical pseudopotential calculations on the band structure of \ce{GaBi} and \ce{GaBiAs} can be found in Refs.\ \cite{janotti_theoretical_2002,zhang_similar_2005,alberi_valence-band_2007,deng_band_2010,bannow_configuration_2016,bannow_valence_2017,pashartis_localization_2017}.  Tight-binding investigation of dilute Bi concentrations in \ce{GaBi_{x}As_{1-x}} have been published by O'Reilly et al.\cite{usman_tight-binding_2011,usman_impact_2013}  In Sec.~\ref{subsec:intro-gabias_bbs-periodic}, we compare our model with these results.  For small lattices with periodic boundaries, we verify the existence of a Bi ``defect state" within the valence band of \ce{GaBi_{x}As_{1-x}} and BAC effects with band edge bowing stronger than traditional alloys \cite{usman_tight-binding_2011}.  In Sec.~\ref{subsec:intro-gabias_bbs-random_cell}, we investigate the electronic structure of large \ce{GaBi_{x}As_{1-x}} lattices of roughly 700000 atoms.  Calculation for the larger structures are carried out with hard infinite boundaries instead of a periodic boundary condition.  The use of large lattice structures allows us to better understand alloy effects at the scale needed for QD structures, as well as to investigate effects caused by random alloy configurations.

\subsection{Band structure of \ce{GaBi_{x}As_{1-x}}: Periodic model} \label{subsec:intro-gabias_bbs-periodic}

\paragraph*{}
To understand the band structure of GaBiAs random alloys, we begin with periodic models of the alloy.  First, we start with simple periodic eight-atom cells of \ce{GaAs}, \ce{GaBi}, and \ce{GaBi_{0.25}As_{0.75}}.  We calculate the folded band structure by diagonalizing the eight-atom TB Hamiltonian.  Then, we expand our model to larger unit cells to explore Bi clustering effects.  More details on the construction of the periodic TB Hamiltonian can be found in Appendix~\ref{appx:gabias}.

\begin{figure}[ht]
  \centering
  \subfloat[GaAs]{
    \includegraphics[width=0.4\textwidth]{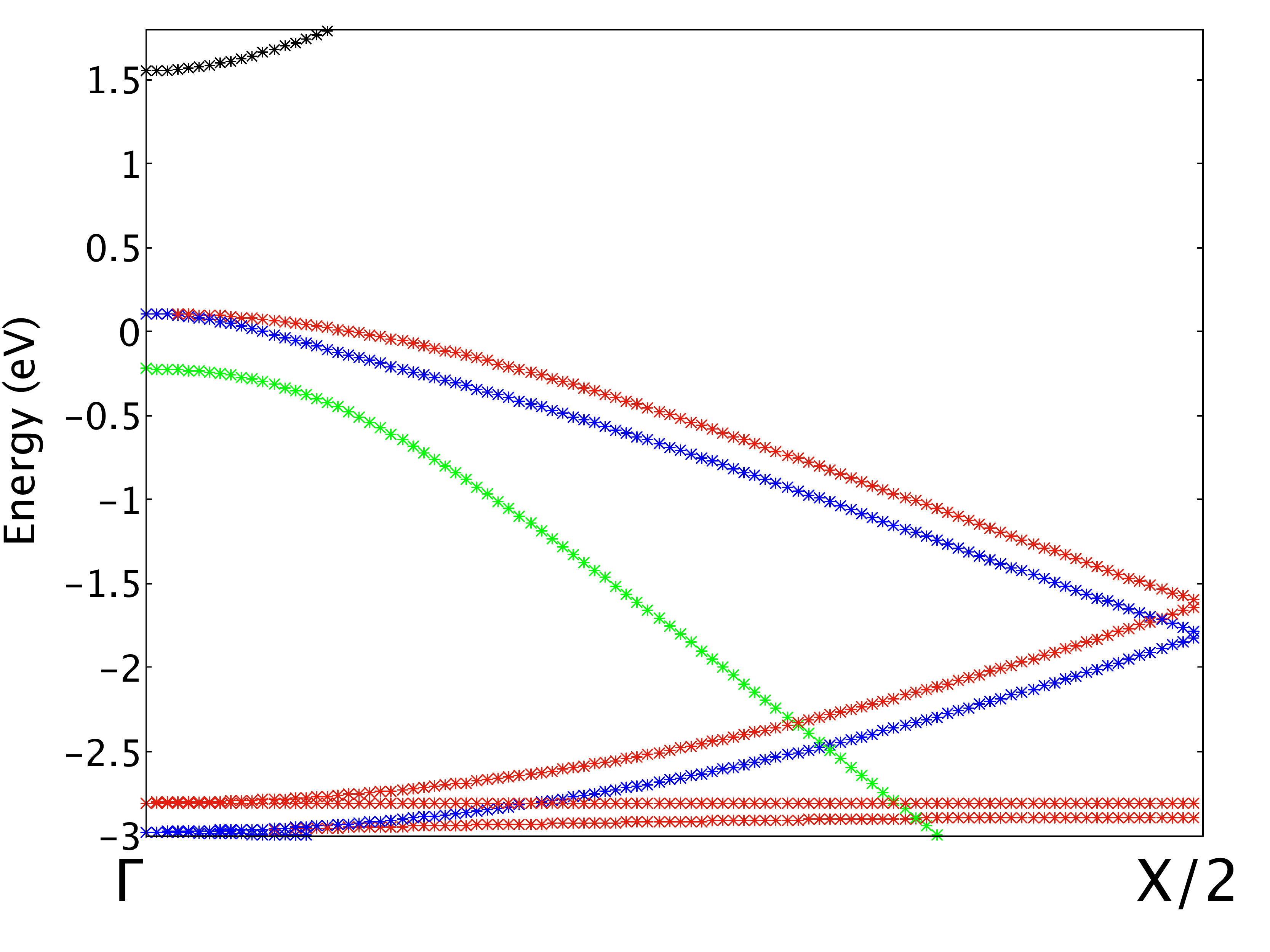}
    \label{fig-sub:bbs_gaas}
  }
  \hspace{0em}
  \subfloat[GaBi]{
    \includegraphics[width=0.4\textwidth]{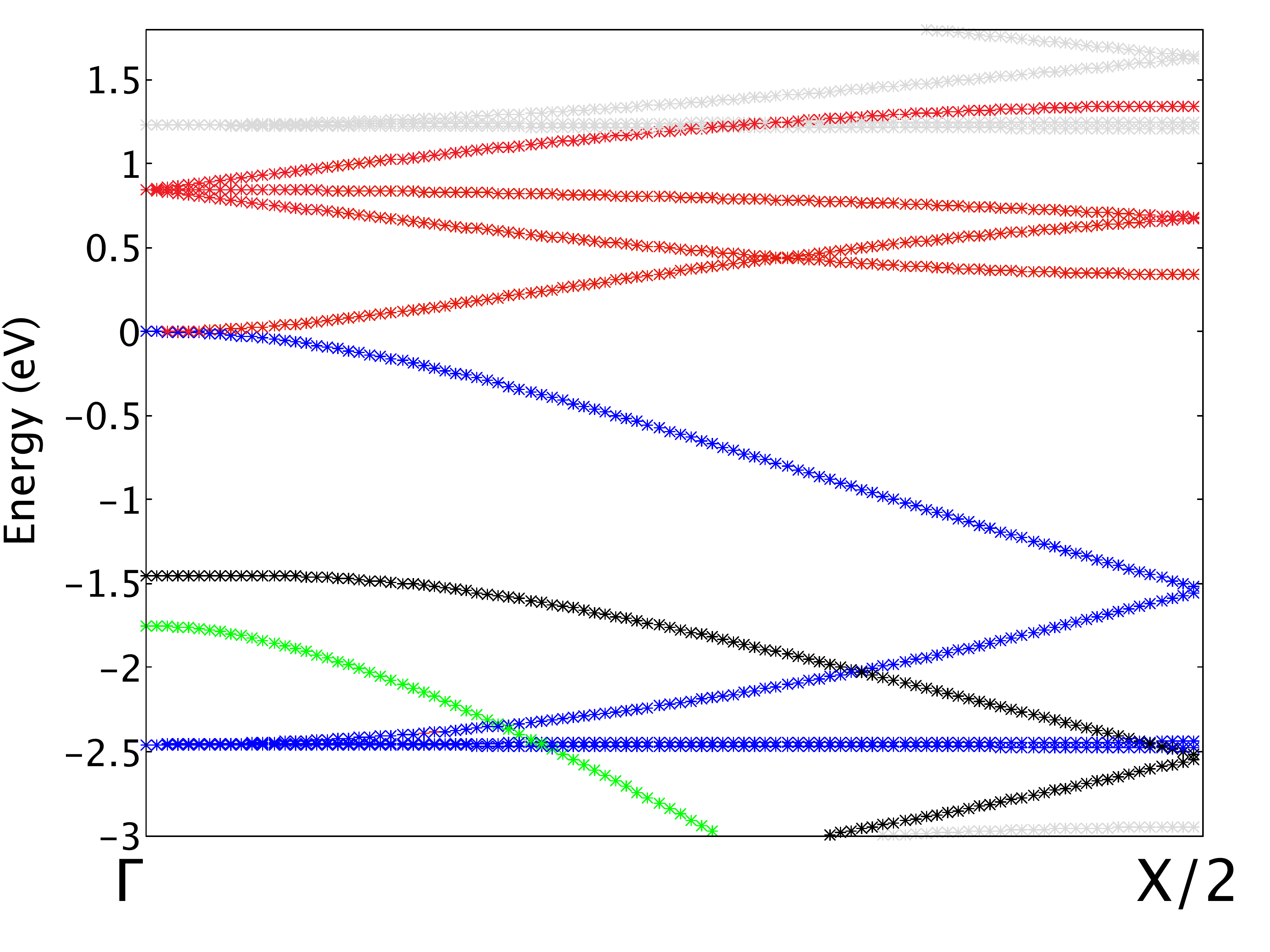}
    \label{fig-sub:bbs_gabi}
  }
  \caption[Band structure of \ce{GaAs} and \ce{GaBi}.]{\small (Color online) Folded band structure of (a) \ce{GaAs} and (b) \ce{GaBi}, from $ \Gamma $ to $ X $.  For an eight-atom cell, the band structure is twice folded.  Both graphs have the VBE set to 0 (\si{\electronvolt}) for the respective material.  (Black: conduction band edge; red: heavy holes; blue: light hole; green: spin-orbit split off; grey: other bands)}
  \label{fig:bbs_gaas_gabi}
\end{figure}

\paragraph*{}
For pure GaAs and GaBi, the folded band structures along $ k $ from $ \Gamma $ to $ X/2 $ are shown in Fig.~\ref{fig:bbs_gaas_gabi}.  GaAs shows the typical behavior of a III--V semiconductor, with a calculated bandgap of \SI{1.44}{~\electronvolt} and an SO split-off \SI{0.33}{~\electronvolt} below the VBE.  GaBi, on the other hand, has an inverted band structure.  The consequence, seen in Fig.~\ref{fig-sub:bbs_gabi}, is that the CBE is at \SI{-1.45}{~\electronvolt}, \emph{below} the VBE (set to 0).  Also, the heavy-hole bands increase in energy away from the zone center while the electron and light-hole bands decrease in energy away from the zone center.  Finally, the SO band for GaBi lies slightly below the conduction band, at \SI{-1.75}{~\electronvolt}.

\begin{figure}[ht]
  \centering
  \includegraphics[width=0.45\textwidth]{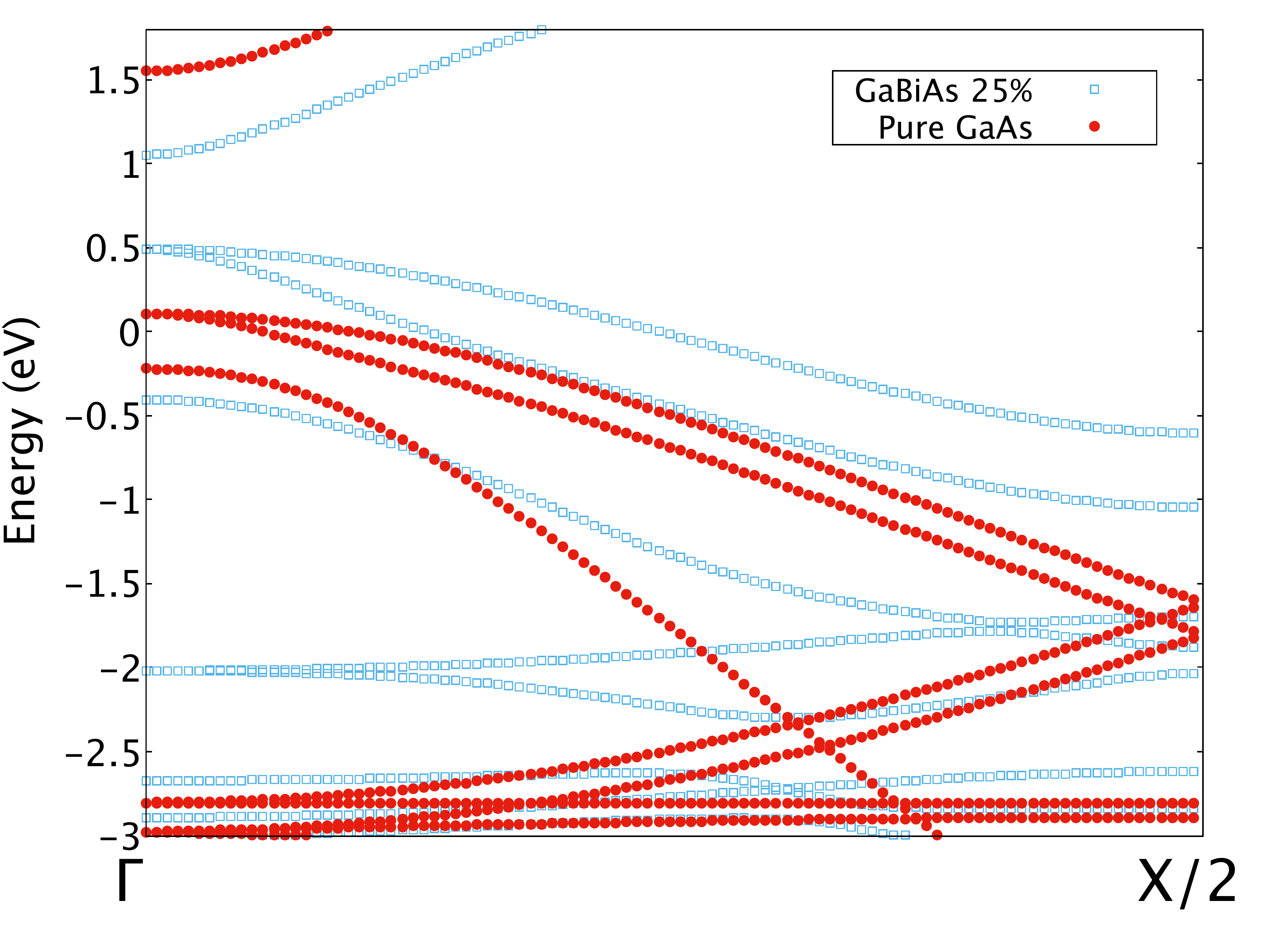}
    \caption[Band structure of \ce{GaBi_{0.25}As_{0.75}}.]{\small (Color online)  Band structure of \ce{GaBi_{0.25}As_{0.75}} (blue), overlaid on top of pure GaAs (red).  One can clearly see the Bi defect state at \SI{-2}{~\electronvolt}.  Placement of Bi atom at the center of the cell breaks the symmetry, thus, the \ce{GaBi_{0.25}As_{0.75}} band structure is not folded.}
  \label{fig:bbs_gabias}
\end{figure}

\paragraph*{}
Next, we take an eight-atom unit cell of pure GaAs and replace the center As atom with a Bi atom, simulating a 25\% Bi \ce{GaBiAs} alloy where the Bi atoms are periodically spaced in the lattice.  Again, we diagonalize this Hamiltonian for a set of $ k $ values between the zone center ($ \Gamma $) and the zone boundary along X.  The resulting band structure, overlaid with the previous results for pure GaAs, are shown in Fig.~\ref{fig:bbs_gabias}.  

\paragraph*{}
There are two primary features of alloying.  First, the alloying reduces the bandgap energy and increases the SO split-off energy.  Second, a Bi band exists beneath the VBE, which is unique to to only a handful of alloy compositions.  The alloy behaves as a non-traditional alloy as it is a mixture of a semiconductor and a semi-metal with drastically different band structures.  The reduction in bandgap energy can be explained under the BAC model, wherein the introduction of a Bi band below the GaAs VBE pushes the alloy VBE up in energy more than traditional alloying \cite{lindsay_theory_1999,alberi_valence-band_2007,usman_tight-binding_2011}.  For low alloy concentrations, the Bi energies lay slightly below the VBE of GaAs, resulting in a BAC effect \cite{usman_tight-binding_2011, deng_band_2010}.  The BAC effect disappears at higher concentrations, because band broadening occurs and the Bi band starts to overlap with the VBE \cite{deng_band_2010}.  In Fig.~\ref{fig:bbs_gabias}, there is a single ``Bi defect state" at roughly \SI{2}{~\electronvolt} below the VBE.  There is limited Bi band broadening due to the presence of only a single Bi atom in the unit cell, therefore, we have a clearly visible defect state.  The existence of this state supports the BAC model \cite{usman_tight-binding_2011}.

\paragraph*{}
Studies have suggested that Bi atoms in \ce{GaBi_xAs_{1-x}} tend to form clusters \cite{usman_tight-binding_2011,bannow_configuration_2016,bannow_valence_2017,laukkanen_local_2017}.  We explore clustering by selectively substituting additional As atoms in the original unit cell with Bi. To simulate change in concentration, the size of the unit cell is expanded by adding GaAs equally to all sides.  Since all the Bi atoms are located in the original eight-atom cell, the periodic repetition of the structure simulates equally spaced ``Bi clusters.'' O'Reilly \textit{et al.\ }varied concentration \cite{usman_tight-binding_2011} similarly in studying a single Bi atom isolated in GaAs.

\begin{figure}[ht]
  \centering
  \includegraphics[width=0.45\textwidth]{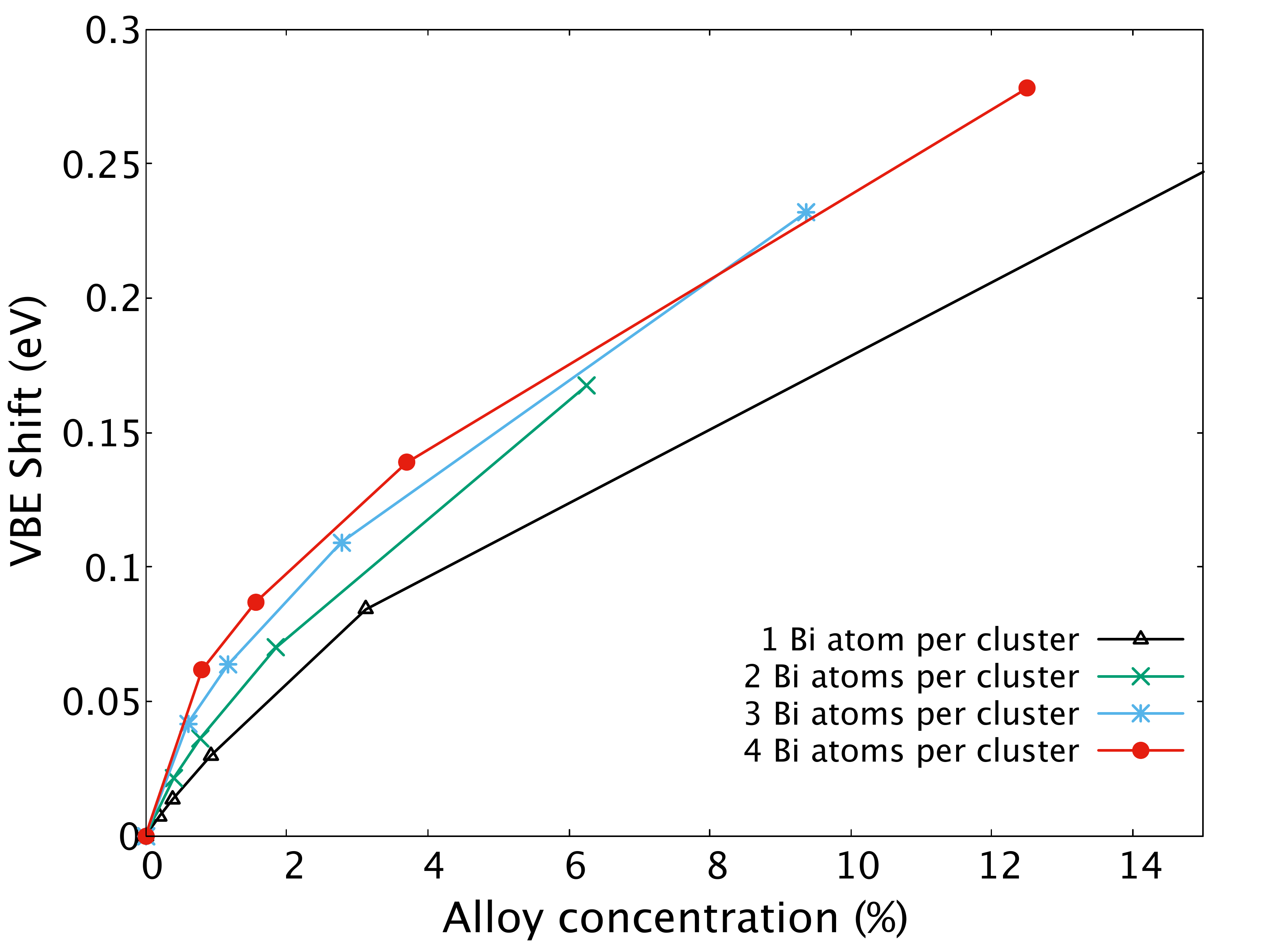}
    \caption[VBE shift of clustered \ce{GaBi_{x}As_{1-x}}.]{\small (Color online) VBE shift above GaAs for \ce{GaBi_{x}As_{1-x}} of different concentrations of Bi.  Different curves represent a different number of Bi atoms in a cluster (located in the center of the periodic cell).  Change in Bi concentration along a curve is created by varying the cell size (the distance between clusters).}
  \label{fig:cluster_gabias}
\end{figure}

\paragraph*{}
The resulting VBE shifts at $ \Gamma $ for the different cell sizes and Bi concentrations are shown in Fig.~\ref{fig:cluster_gabias}.  Each connected curve represents a given number of Bi atoms per cluster; each data point along a given curve corresponds to a different unit cell size, with larger Bi concentrations for smaller cell sizes.  Following each curve, we find the smaller the cell size, the higher the concentration, and the more the VBE is shifted.  Comparing curves, we see that higher Bi atom counts per cluster leads to greater rise in energy for the VBE, despite identical Bi concentration.  This implies that alloying effects are increased by clustering, consistent with O'Reilly \cite{usman_tight-binding_2011,usman_impact_2013}.  For Bi concentrations in the 4\% to 10\% range, clustering leads to shifts ($\sim$\SI{50}{~\milli\electronvolt}) that are a sizable  fraction of the actual VBE shifts. 

\subsection{Band structure of \ce{GaBi_{x}As_{1-x}}: Random alloy} \label{subsec:intro-gabias_bbs-random_cell}

\begin{figure}[ht]
  \centering
  \includegraphics[width=0.45\textwidth]{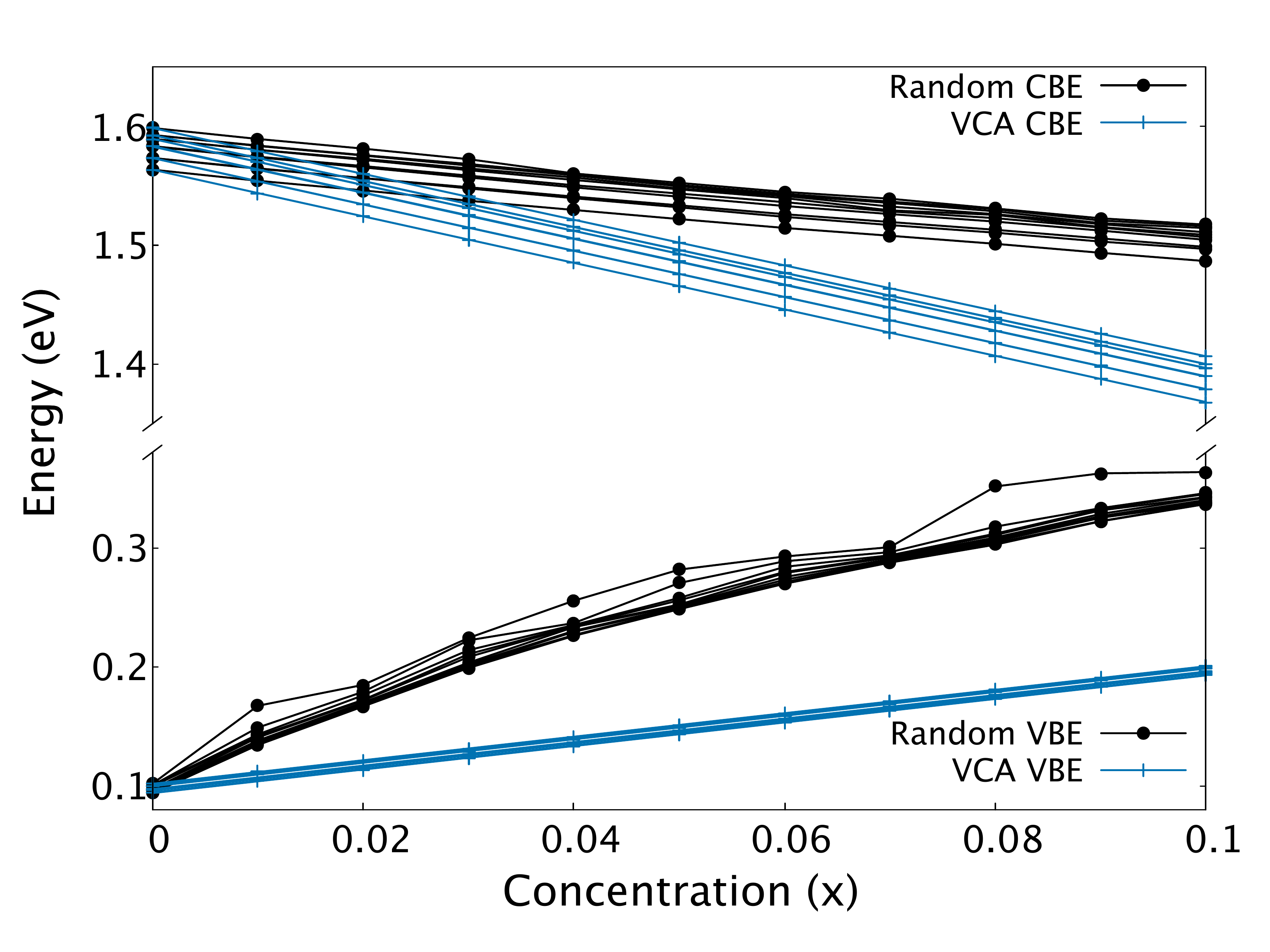}
    \caption[Band edge of bluk \ce{GaBi_xAs_{1-x}}.]{\small (Color online) Energy levels of \ce{GaBi_xAs_{1-x}} plotted against Bi concentration (x), for a random alloy approach (black) and a VCA model (blue).  The zero on the energy scale here is set to the VBE for \ce{GaAs}.  The initial (0\% Bi) deviation from a perfect \SI{0}{~\electronvolt} GaAs VBE is caused by spin-orbit effects not included by Vogl \cite{vogl_semi-empirical_1983}.}
  \label{fig:cell_be}
\end{figure}

\paragraph*{}
To better model actual QDMs, we must consider random alloys of GaBiAs.  We take a $ 144a \times 144a \times 45a $ box of pure GaAs, and use a pseudo-random number generator to select As atoms to replace with Bi.  The large lattice is then relaxed using a VFF method to account for strain due to Bi in the alloyed system.  A TB Hamiltonian is generated from a $ 50a \times 50a \times 34a $ cutout of the relaxed lattice.  The Hamiltonian is diagonalized to find the states closest to the VBE and CBE.  A hard boundary (non-periodic) is imposed on both the VFF relaxation and the Hamiltonian diagonalization.  The results are shown as the black curves in Fig.~\ref{fig:cell_be}.

\paragraph*{}
Also shown in Fig.~\ref{fig:cell_be} are results from using a virtual crystal approximation (VCA) to model alloying.  In the VCA model, the alloy is treated as a regular lattice of GaX, where the VFF and TB parameters for X are the concentration-dependent averages between parameters for As and Bi.  The VCA energies have a linear dependence on alloy concentration.  Although both the VCA model and the random alloy approach show a bandgap reduction of roughly \SI[per-mode=symbol]{30}{\milli\electronvolt\per\percent}, only the random alloy case correctly captures bowing of the VBE.  In addition, as indicated in Table~\ref{table:cell_be}, the VBE shift obtained with a VCA model is smaller than the prediction of O'Reilly's \cite{usman_tight-binding_2011} and experimental results \cite{yoshida_temperature_2003,fluegel_giant_2006,alberi_valence-band_2007,laukkanen_local_2017}.  The failures of the VCA model show the necessity of including the individual Bi atoms and reaffirm the importance of the BAC model.  The results for our random alloy model do differ slightly from O'Reilly's but can be explained by a difference in strain models.  Our results show that an atomistic treatment of Bi is needed to correctly predict the properties of \ce{GaBi_{x}As_{1-x}}, which would be a random alloy in real materials.

\begin{table}[ht]
  \centering
  \caption[Band-edge shift for \ce{GaBi_{x}As_{1-x}}.]{\small Band edge shift per alloy percentage (slope in Fig.~\ref{fig:cell_be}) for \ce{GaBi_{x}As_{1-x}}.  Two values taken, one between 0\% and 1\%, the other between 9\% and 10\%.  Values from O'Reilly \textit{et al.}, where the system was a smaller random alloy with periodic boundaries, are shown for comparison \cite{usman_tight-binding_2011}.  The gap energy, $ E_g $, decreases by the combined amount of the VBE and CBE shifts.  All units in \si[per-mode=symbol]{\milli\electronvolt\per\percent}.} \label{table:cell_be}
  \begin{ruledtabular}
  \begin{tabular}{l|ddd}
    & \multicolumn{1}{l}{CBE} & \multicolumn{1}{l}{VBE} & \multicolumn{1}{c}{$ E_g $} \\
    \colrule
    VCA (1\%) 	&  -19	&  9.8	&  -28.8 \\
    VCA (10\%) 	&  -19	&  9.8	&  -28.8 \\
    Alloy (1\%) 	&  -9.2	&  66	& -75.2 \\ 
    Alloy (10\%) 	&  -5.4	&  11	& -16.4 \\
    \colrule
    Periodic \cite{usman_tight-binding_2011} (2\%) 	&  -28	&  53	& -81 \\
    Periodic \cite{usman_tight-binding_2011} (12\%)	&  -28	&  30	& -58 \\
  \end{tabular}
  \end{ruledtabular}
\end{table}

\begin{figure}[ht]
  \centering
  \includegraphics[width=0.45\textwidth]{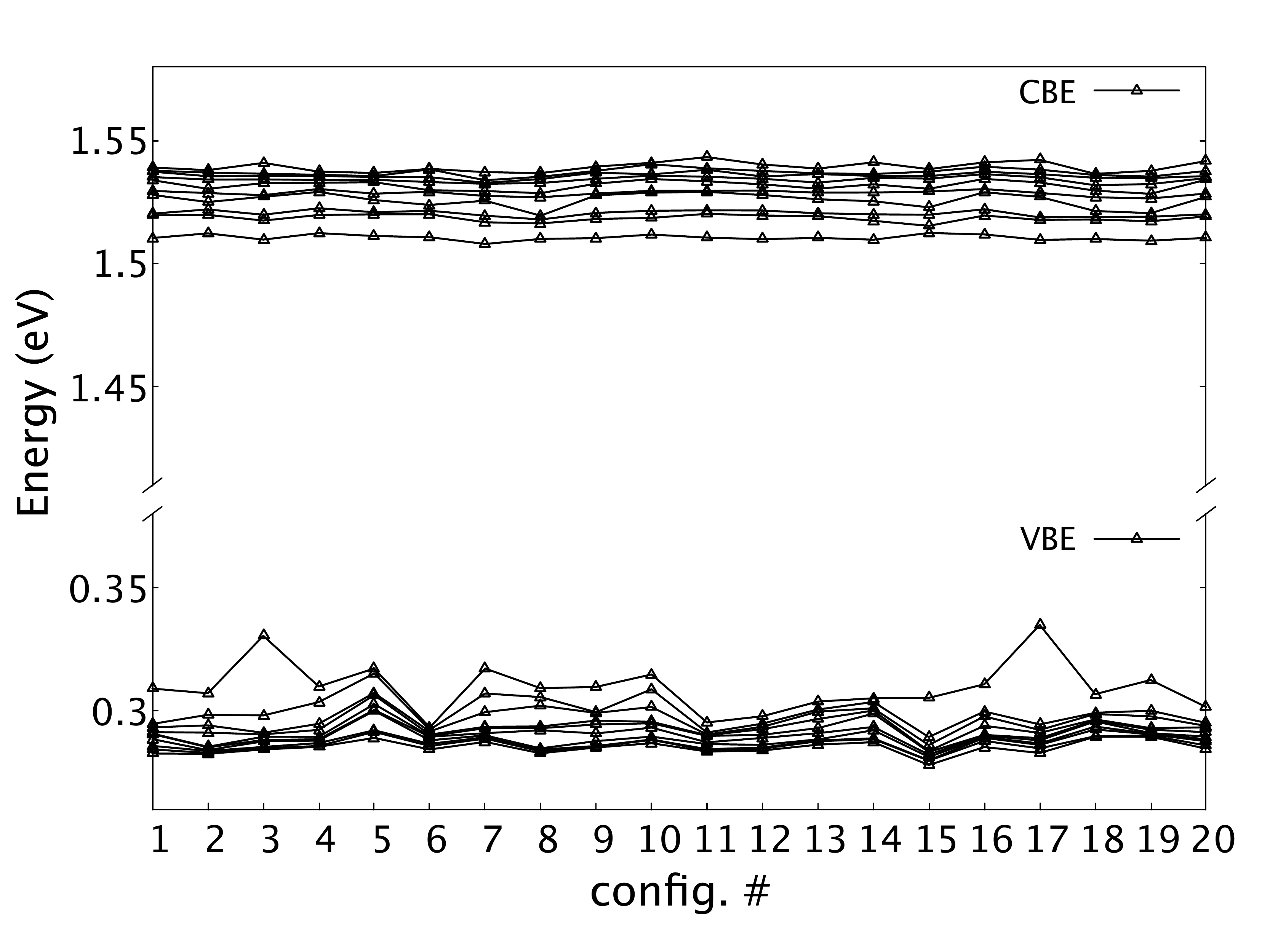}
    \caption[Configurational spread of band edges of bulk \ce{GaBi_{0.07}As_{0.93}}.]{\small Band edge energies for various random alloy configurations with a \ce{GaBi_{0.07}As_{0.93}} cell of nearly 700,000 atoms. The zero on the energy scale here is set to the VBE for GaAs.  Data points are connected via curves solely for better visualization.}
  \label{fig:cell_be-config}
\end{figure}

\paragraph*{}
With the random alloy model, we also study the spread of energies stemming from variations in alloy configuration.  These variations are not accounted for by the VCA nor in a periodic model.  For the case of 7\% Bi, 40 different alloy configurations (placement of Bi atoms) are generated by the pseudorandom number generator.  Each configuration is relaxed and diagonalized independently.  The resulting energies closest to the band edges for each configuration are shown in Fig.~\ref{fig:cell_be-config}.  We see a \SI{50}{~\milli\electronvolt} spread for the VBE and a \SI{10}{~\milli\electronvolt} spread for the CBE.  In conjunction with Fig.~\ref{fig:cell_be} and Table~\ref{table:cell_be}, we see that the configurational variances accounts for a large portion of the difference between our predicted energies and O'Reilly's.  The remaining discrepancies between the random alloy results and O'Reilly's paper can be attributed to the difference between non-periodic and periodic boundary conditions, as well as the fact that we have a different model for strain relaxation.  A later paper by O'Reilly \textit{et al.\ }\cite{usman_impact_2013}, where \ce{GaBi_xAs_{1-x}} was placed on top of a GaAs substrate (consistent with our strain model), shows band edge shifts in agreement with our random alloy results.

\paragraph*{}
In conclusion, for a periodic system, we have reaffirmed the existence of a Bi state below the band edge which pushes the VBE up through BAC.  Stronger alloy effects emerge from more densely clustered Bi.  For large nonperiodic lattices, the VBE shift of the random alloy model is consistent with literature, while the VCA model is not consistent with random alloy results.  Furthermore, the VBE of the random alloy model shows signs of bowing, an effect of BAC \cite{alberi_valence-band_2007,usman_tight-binding_2011}, where the VCA model failed to show signs of BAC.  Additionally, the spread in energies at the VBE for the random alloy case is a combined effect of configurational dependence and strain \cite{usman_impact_2013}.  While the VB is sensitive to both strain and configuration due to the BAC, the lack of BAC in the CB removes most of the configurational dependence from the CBE.  These results show that an atomistic treatment is necessary to correctly model the properties of \ce{GaBi_{x}As_{1-x}}.  Most importantly, an atomistic treatment is needed to capture the VBE shift, which significantly determines effects from changes in the barrier in a QDM.


\section{Quantum Dot Molecules with Alloyed Barrier} \label{sec:qdm_gabias}

\paragraph*{}
To exploit GaBiAs barriers to manipulate hole spin physics in InAs QDMs, we must understand how GaBiAs barriers modify QD energy levels, how the results depend on alloy concentration and configuration, how the strain from the Bi influences the results, and how the defect levels introduced by Bi change the energies.  These issues, discussed in this section, provide the foundation to understand tunnel coupling and spin-mixing between QDs.  The use of GaBiAs barriers for the enhancement of the spin-mixed QD states needed in the qubit protocol of Econumou \textit{et al.\ }\cite{economou_scalable_2012} will be discussed in the subsequent paper.

\subsection{\ce{InAs} QDM states with \ce{GaBi_{x}As_{1-x}} interdot barriers} \label{subsec:gabias-dot}

\begin{figure}[ht]
  \centering
  \includegraphics[width=0.45\textwidth]{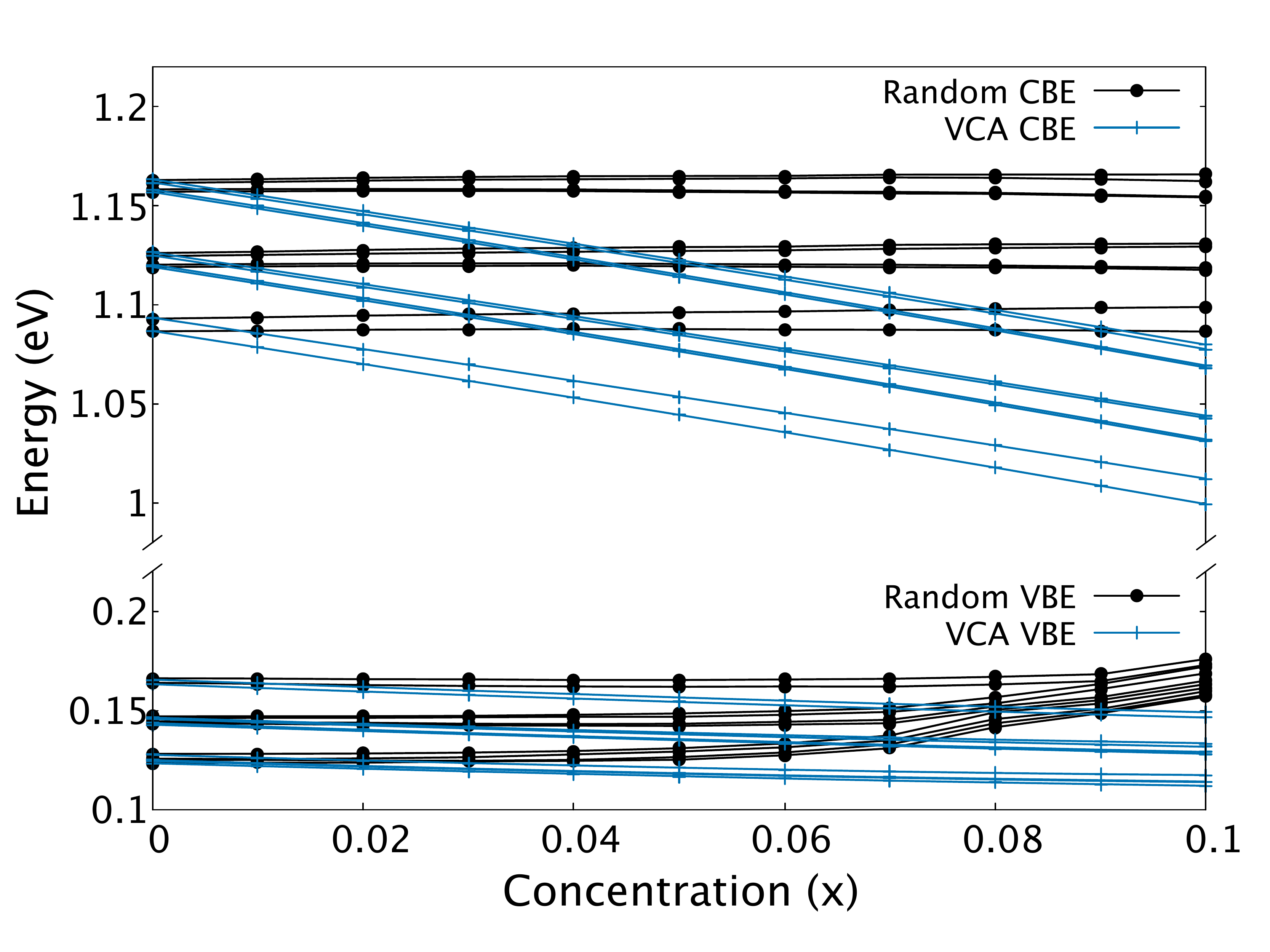}
    \caption[Energy levels of an \ce{InAs} QDM with full \ce{GaBi_xAs_{1-x}} barrier.]{\small (Color online) Energy levels of an \ce{InAs} QDM with full \ce{GaBi_xAs_{1-x}} barrier, calculated with a random alloy approach (black) and the VCA (blue).  Zero on the energy scale here and in remaining figures is set to the VBE for \ce{InAs}, which is \SI{0.2}{~\electronvolt} above the \ce{GaAs} VBE.}
  \label{fig:dot_be-random}
\end{figure}

\paragraph*{}
We first consider a QDM with GaBiAs everywhere between the InAs QDs and wetting layers [as shown in Fig.~\ref{fig-sub:system_geo-full}].  Energies obtained from the VCA model are shown in Fig.~\ref{fig:dot_be-random} (blue), in comparison with results obtained for a random alloy configuration (black).  From here on, we shall make the distinction between QD hole/electron states and the VB/CB states.  The nomenclature ``holes'' and ``electrons'' shall be reserved for the states confined within the QDs, whereas VB and CB shall refer generally to electron states below or above the bulk bandgaps.  All energy plots will show conduction electron and valence electron energies.  Due to convention, an increase in hole state energy is a decrease in valence electron energy, and vice versa.

\paragraph*{}
As shown in Fig.~\ref{fig:dot_be-random}, electron states confined to the dots are weakly affected by the presence of Bi, due to the relatively small \ce{GaBi_xAs_{1-x}} CBE energy shift.  The larger energy difference between the dot electron states and the barrier \ce{GaBi_xAs_{1-x}} CBE also weakens the barrier alloy effects as compared to the holes.  From Fig.~\ref{fig:cell_be}, we see that the CBE for bulk \ce{GaBi_xAs_{1-x}} is about \SI{0.3}{~\electronvolt} above that of the InAs QD electron states (after accounting for the VB offset between GaAs and InAs).  The atomistic treatment of Bi accounts for local Bi effects, without distorting the electron structure globally, as would happen in the VCA.

\paragraph*{}
The hole state energies in Fig.~\ref{fig:dot_be-random} are initially flat, weakly dependent on Bi concentration, because the states closest to the gap are well confined to the \ce{InAs} dots.  In the random alloy model, this breaks down at higher concentrations of Bi, when the GaBiAs barrier VBE is pushed up into the same energy region as the confined QD hole states.  This results in a breakdown of confinement at 9\% to 10\% Bi.

\begin{figure}[ht]
  \centering
  \includegraphics[width=0.45\textwidth]{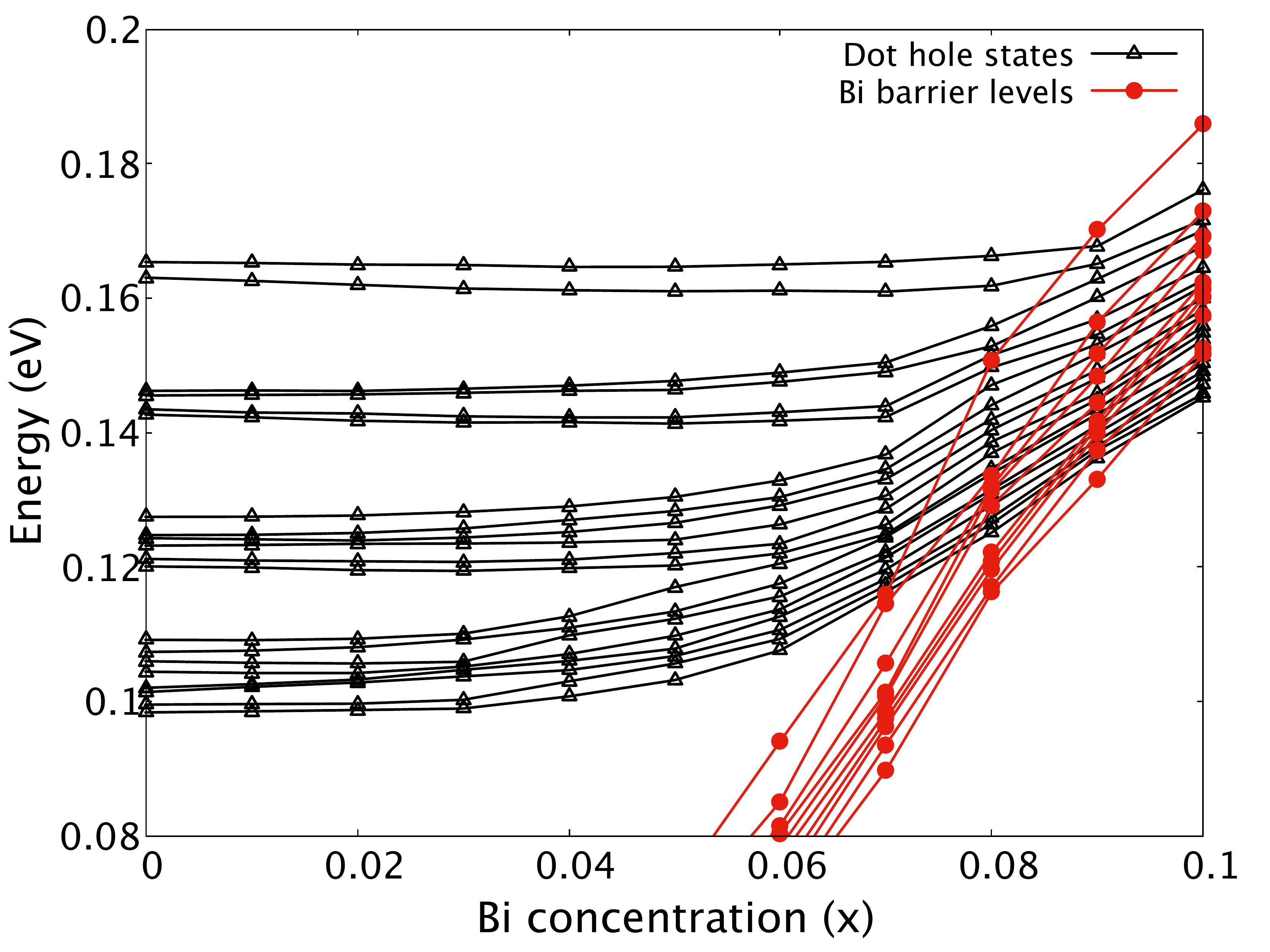}
    \caption[Comparison of barrier valence band energy with QD hole state energy.]{\small (Color online) Comparison of GaBiAs barrier valence band energy levels (red) with random alloy \ce{InAs} QD hole state energy levels (black).  Bi barrier energy levels were calculated as a layer of \ce{GaBi_xAs_{1-x}} in a larger lattice of \ce{GaAs}.}
  \label{fig:spacing-7-Bi_bands}
\end{figure}

\paragraph*{}
The interaction between the dot hole states and the barrier VBE is supported by the fact that the hole states are close in energy to the VB of bulk \ce{GaBi_xAs_{1-x}} at $x \approx 7$\%.  We see the distortion of the dot-hole-state energy levels as the concentration of Bi in the barrier surpasses this percentage.  To further illustrate this, we calculate the energies with the two InAs dots and wetting layers replaced with GaAs, leaving just the \ce{GaBi_xAs_{1-x}} sandwiched between GaAs.  The energies for this \ce{GaBi_xAs_{1-x}} layer are compared with the energies for the InAs dots in Fig.~\ref{fig:spacing-7-Bi_bands}.  This better illustrates the distortion of the dot energy levels with respect to the encroaching Bi band energy levels.  This also indicates that, past 8\% Bi, the valence states need not be confined solely to the QDs.  Wave-function plots confirm that, for 10\% Bi, significant hole probability is found in the Bi barrier region rather than in the InAs dots.

\begin{figure}[ht]
  \centering
  \includegraphics[width=0.45\textwidth]{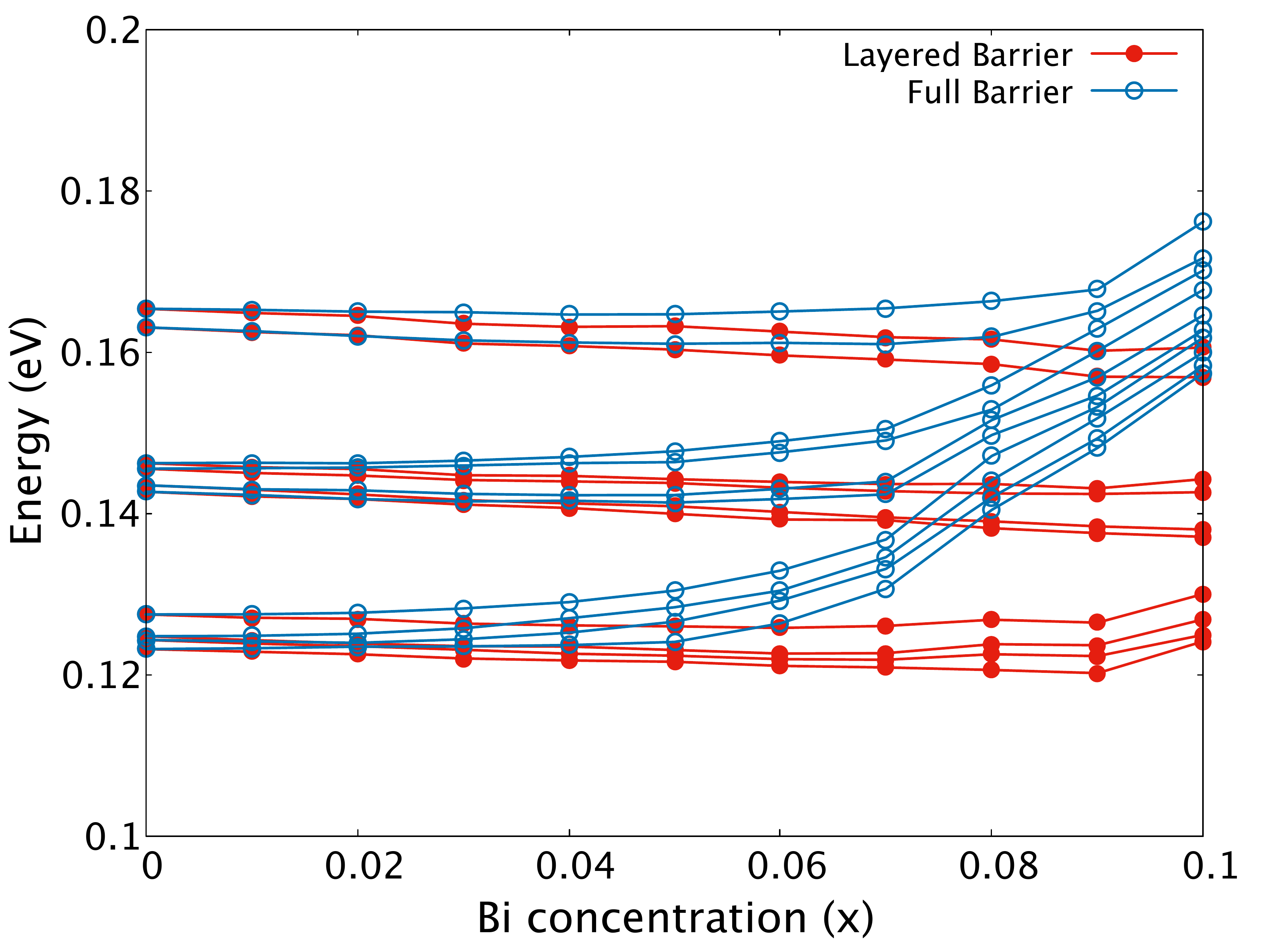}
    \caption[Energy levels of a \ce{InAs} QDM with a partial \ce{GaBi_xAs_{1-x}} barrier.]{\small (Color online) valence electron energies of an \ce{InAs} QDM with the interdot region containing a $ 4a $ layer of \ce{GaBi_xAs_{1-x}} (red) compared to a QDM where the entire interdot region is \ce{GaBi_xAs_{1-x}} (blue).}
  \label{fig:dot_vbe-layer}
\end{figure}

\paragraph*{}
We also consider the case where only a $ 4a $ layer within the interdot region is alloyed [see Fig.~\ref{fig-sub:system_geo-part}].  This shows how the thickness of the alloy barrier region can influence the QDM states.  The resulting hole energies, overlaid with the dot hole energies with the full GaBiAs barrier, are shown in Fig.~\ref{fig:dot_vbe-layer}.  There are two key things to note.  First, for the layered GaBiAs barrier, the upward shift in valence energy doesn't occur until a higher concentration of Bi, indicating that the QD hole states are less perturbed by the alloy region, despite having the same concentration of Bi within the alloy layer.  Second, for the full GaBiAs barrier there is a larger energy difference between equivalent top and bottom dot states (i.e.,\ $E_{v1} - E_{v2}$, $E_{v3,v4} - E_{v5,v6}$, etc.), than for the layered barrier.  This is because, in the former case, there is Bi present above and around the sides of the bottom dot, whereas Bi is only present below the top dot.  This geometry of the GaBiAs layer breaks the symmetry between the top and bottom dots.  On the other hand, the $ 4a $ layer of GaBiAs sandwiched  between GaAs is symmetric about the two dots.  We will later see in Sec.~\ref{subsec:gabias-strain} that the increased energy difference is a strain induced effect.

\subsection{Alloy configuration dependence} \label{subsec:gabias-config}

\begin{figure}[ht]
    \includegraphics[width=0.45\textwidth]{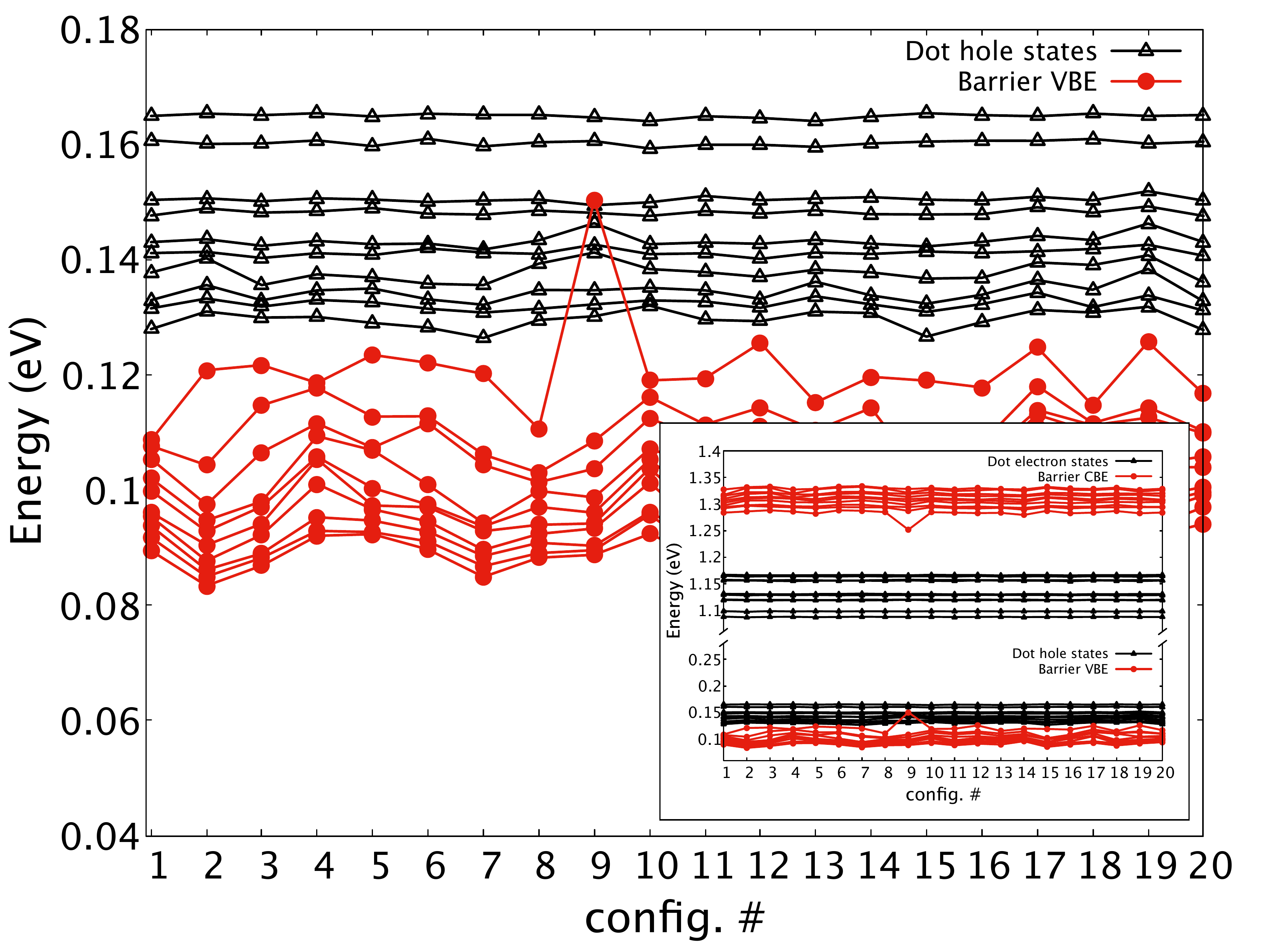}
	\caption[\ce{InAs} QDM states with various barrier \ce{GaBi_{0.07}As_{0.93}} alloy concentrations.]{\small (Color online) Valence edge states for various random alloy configurations of a $ 10a $ layer of \ce{GaBi_{0.07}As_{0.93}} in \ce{GaAs} (red), and QD states from using the respective alloy configuration in the interdot barrier (black).  Inset shows both valence and conduction band edge states on a rescaled $y$-axis.  Zero on the energy scale is set to the VBE for \ce{InAs}.}
	\label{fig:config_dot}
\end{figure}

\paragraph*{}
To test the effect of alloy configuration, we consider InAs QDMs with 7\% Bi in the GaBiAs barrier.  At this percentage, the GaBiAs VBE overlaps with the hole states in the QDM.  For the bulk \ce{GaBi_xAs_{1-x}} (Fig.~\ref{fig:cell_be-config}), there is a \SI{10}{~\milli\electronvolt} spread in CBE energies and a \SI{50}{~\milli\electronvolt} spread in VBE energies caused by different configurations.  The spread of CBE and VBE energy levels for 40 configurations of a $ 10a $ layer of \ce{GaBi_{0.07}As_{0.93}} in GaAs are shown in Fig.~\ref{fig:config_dot} (red).  The alloy configurations in Fig.~\ref{fig:config_dot} have no relation to those in Fig.~\ref{fig:cell_be-config}.  Figures~\ref{fig:cell_be-config}~and~\ref{fig:config_dot} look similar, with the exception of one case for the CBE.  This is an artifact of a particular alloy configuration and is a very rare occurrence.  Looking at the VBE, it might seem that there is more of a spread for a $ 10a $ layer of GaBiAs in GaAs, compared to bulk GaBiAs.  However, the states are less dense because fewer Bi atoms are included in the computational box.  This is not a configurational effect.  The spread in VBE is still roughly \SI{50}{~\milli\electronvolt}.

\paragraph*{}
The effect of alloy configuration on the QDM electron and hole states is also shown in Fig.~\ref{fig:config_dot}. We take the GaAs lattice with a GaBiAs barrier shown in red in Fig.~\ref{fig:config_dot} and add the InAs dots.  The alloy configuration with and without the dots remains the same for the same configuration numbers.  The resulting dot states are shown in Fig.~\ref{fig:config_dot} (black).  The electron states remain largely unaffected by alloy configuration, as expected.  The configurational change of the \ce{GaBi_{x}As_{1-x}} CBE is too small and too far away to significantly affect the behavior of the dot electron states.  However, the hole states do see the \ce{GaBi_{x}As_{1-x}} VBE, because, for $ x=0.07 $, the alloy VBE energetically overlaps with the hole states.  At 7\% alloy, the ground and first excited QDM hole states retain their structure regardless of alloy configuration; the remaining states are shifted to varying degrees correlated to how close the \ce{GaBi_{0.07}As_{0.93}} VBE is.

\subsection{Strain-induced effects of a \ce{GaBiAs} alloy} \label{subsec:gabias-strain}

\begin{figure}[ht]
  \centering
  \includegraphics[width=0.45\textwidth]{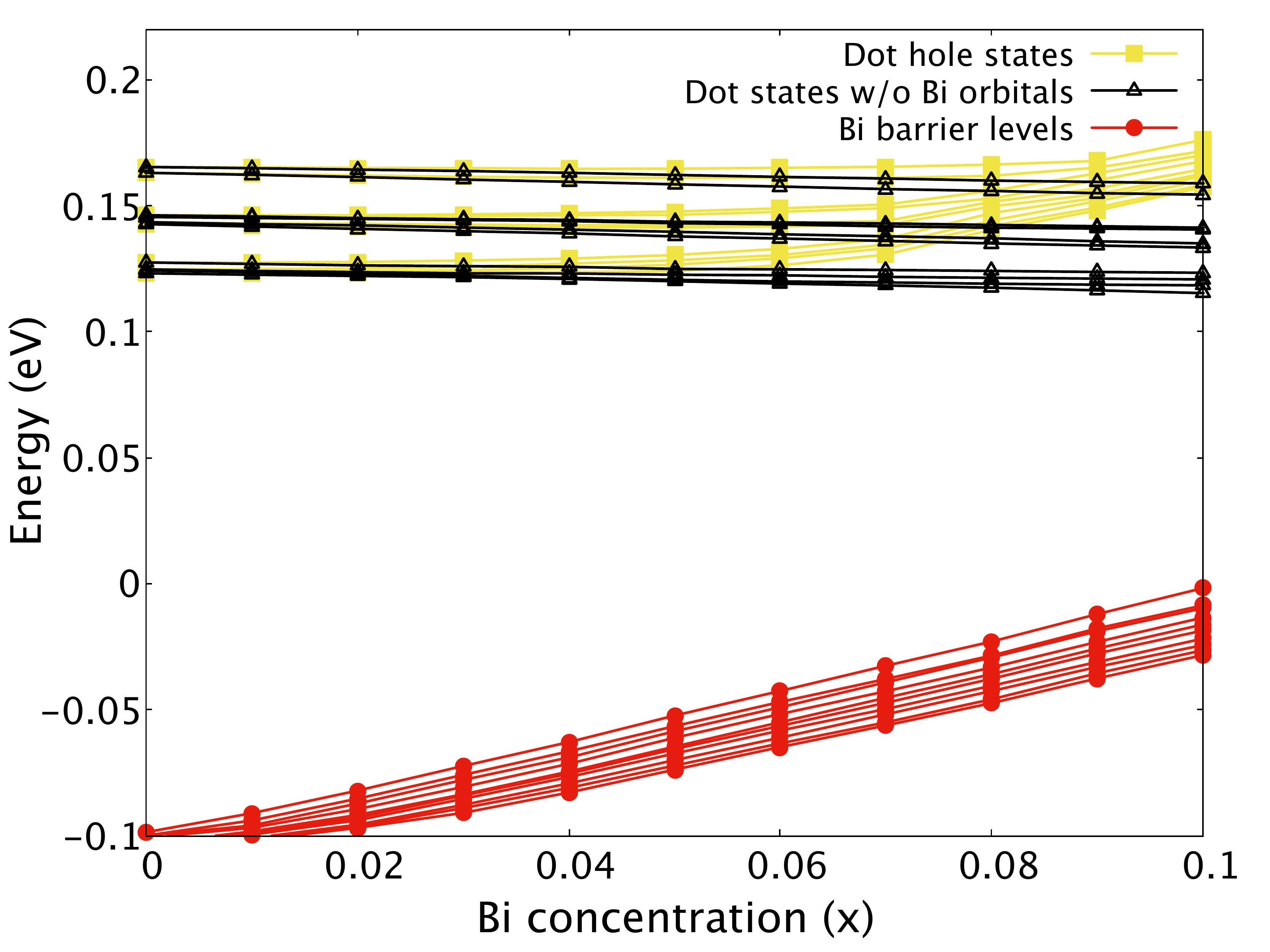}
    \caption[Comparison of barrier valence band energy with QD hole state energy, strain effects only. ]{\small (Color online) Comparison of GaBiAs barrier valence energy levels (red) and \ce{InAs} dot hole state energy levels (black) with all GaBi tight-binding parameters replaced with those of GaAs, keeping the strain profile for the GaBiAs barrier.  Energy levels found with both strain and orbital effects included are shown for comparison (yellow).}
  \label{fig:spacing-7-Bi_bands-not_bi}
\end{figure}

\paragraph*{}
To better understand how the Bi in the barrier perturbs the QDM states, we look at strain and orbital effects of Bi alloying independently.  First, we consider the strain effects caused by the presence of Bi atoms, as there is an 11\% difference in bond length between GaAs and GaBi.  By including the strain from the alloy barrier but using the GaAs TB parameters for GaBi, the effects of strain from Bi are isolated from the effects due to the differences in the Bi and As orbital and hopping energies.  Calculations are done with and without the InAs dots (Fig.~\ref{fig:spacing-7-Bi_bands-not_bi}) to show the dot hole states and the underlying GaBiAs VB.

\paragraph*{}
In Fig.~\ref{fig:spacing-7-Bi_bands-not_bi}, we see a near-linear increase in dot-hole-state energy (i.e., a decrease in valence electron energy) with respect to Bi concentration due to strain.  We also see a steady energy shift of the underlying Bi band due to strain alone.  Although the underlying Bi band does not overlap with the dot hole states in energy, and the Bi is not present where the dot states are spatially, we clearly see that alloy strain has an effect on the QDM hole states.  The strain induced by alloying the barrier has a global effect, not confined to just the alloy region but able to affect the energies of the states in the QDMs.  The strain in the GaBiAs alloy reduces barrier height.  However, the additional strain induced to the QDs counteracts the reduction in barrier height, providing more confinement and increasing the hole energies (see Appendix~\ref{appx:strain_profile}).

\begin{figure}[hb]
	\centering
	\includegraphics[width=0.45\textwidth]{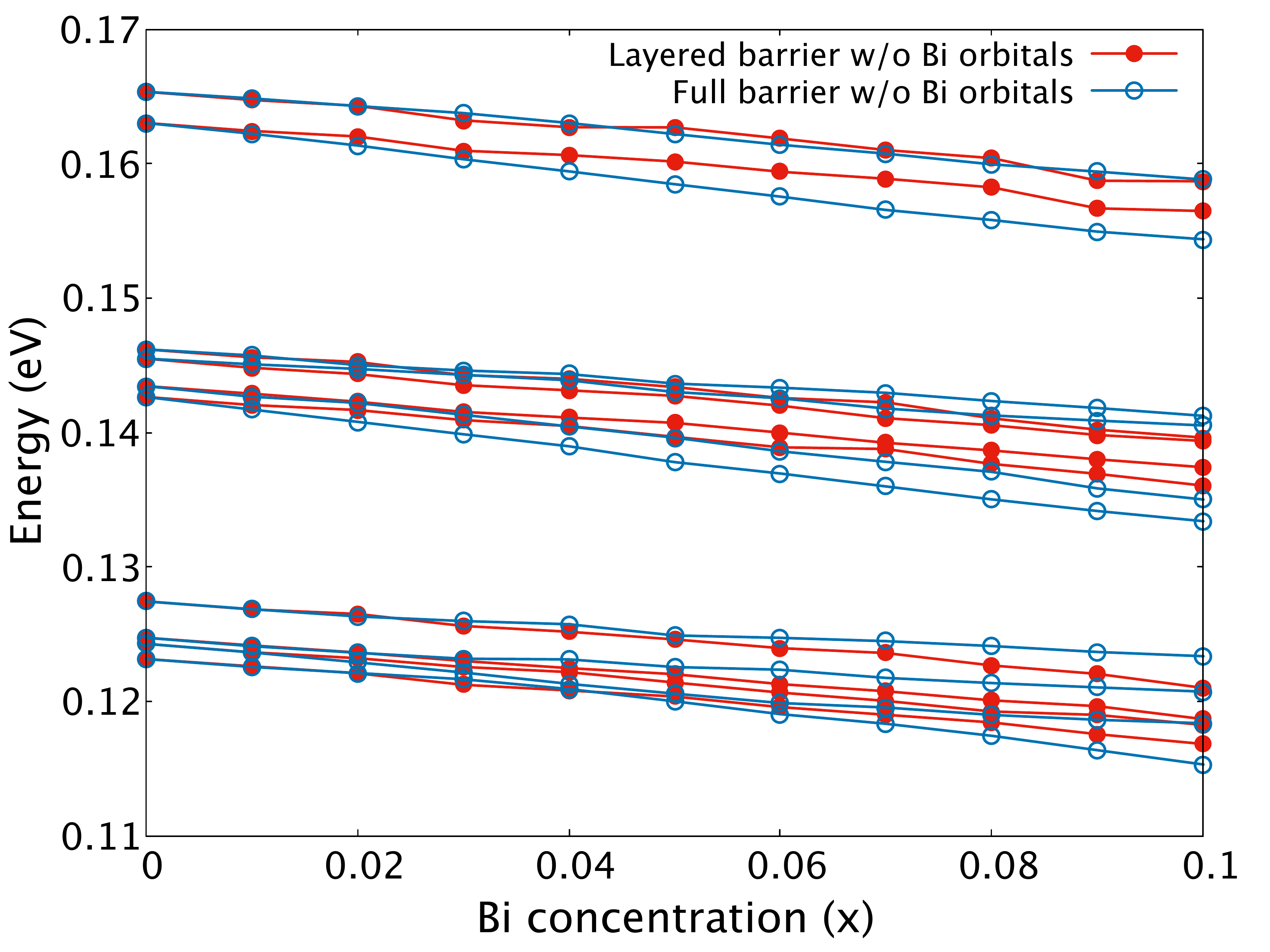}
	\caption[Energy levels of a \ce{InAs} QDM with a partial \ce{GaBi_xAs_{1-x}} barrier, strain effects only.]{\small (Color online) Strain dependence of valence energy levels of an \ce{InAs} QDM, with the interdot region containing a $ 4a $ layer of \ce{GaBi_xAs_{1-x}} (red) or the entire interdot region as \ce{GaBi_xAs_{1-x}} (blue). }
	\label{fig:spacing-7-Bi_bands-not_bi-thickness}
\end{figure}

\paragraph*{}
Figure~\ref{fig:spacing-7-Bi_bands-not_bi-thickness} compares the effect of induced strain for a QDM with a full barrier and a layered barrier.  The fully alloyed barrier provides larger energy splitting between top/bottom dot hole states within a given pair ($E_{v1} - E_{v2}$, $E_{v3,v4} - E_{v5,v6}$, etc.).  This arises from symmetry breaking between the two dots when the sides of the bottom dot are surrounded by Bi.  The similar strain profile in the top and bottom dot for the layered barrier causes energy levels to increase at a similar rate.  Both the fully alloyed barrier and the layered barrier show a monotonic increase in dot hole energy due to strain.  Neither shows a level repulsion from a Bi band.  Slight deviations from a linear dependence show that the strain profile of the system is still sensitive to configurational differences in the alloy, further illustrating the global nature of strain effects on the system.

\subsection{The effect of orbital defects in a \ce{GaBiAs} alloy} \label{subsec:gabias-elec}

\begin{figure}[ht]
  \centering
  \includegraphics[width=0.45\textwidth]{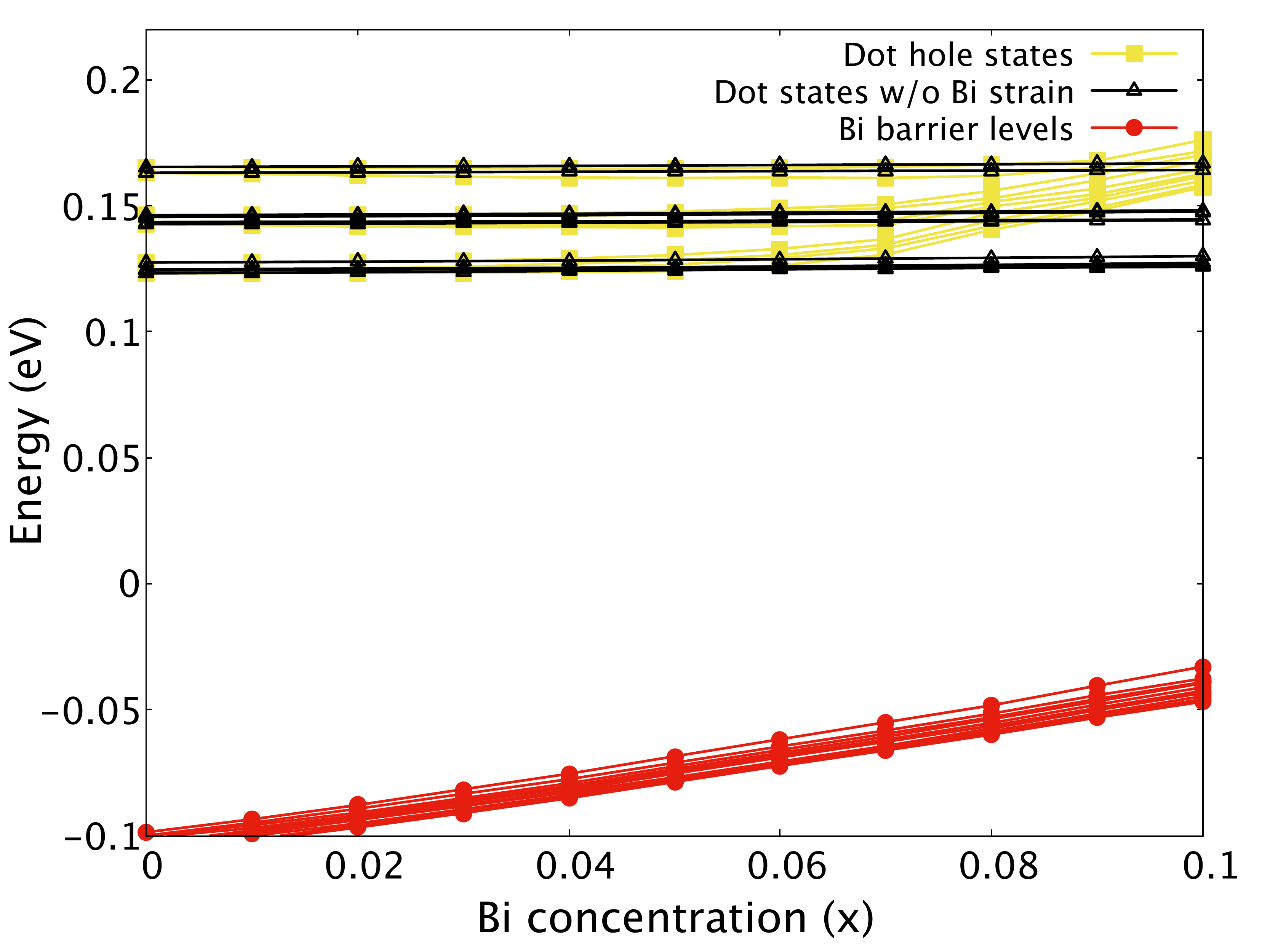}
    \caption[Comparison of barrier valence band energy with QD hole state energy, orbital effects only. ]{\small (Color online)  Comparison of GaBiAs barrier valence energy levels (red) with \ce{InAs} dot-hole-state energy levels (black), using the strain profile of a \ce{GaAs} barrier, effectively leaving only orbital effects of the Bi alloy.  Energy levels found with both strain and orbital effects included are shown for comparison (yellow).}
  \label{fig:spacing-7-Bi_bands-unrelaxed_bi}
\end{figure}

\paragraph*{}
We now look at the effect of the difference between As and Bi orbitals, independent of the strain.  To remove the strain induced by the Bi, we relax the structure with the interdot region as \ce{GaAs}, and use these atomic positions for the GaBiAs barrier.  This new lattice contains the GaBiAs interdot barrier, but uses the ``unrelaxed'' Bi positions.  It is important to note that the system is partially relaxed.  The strain caused by the lattice mismatch between InAs and GaAs is preserved, a crucial aspect of InAs QDMs, regardless of the presence of the alloy.  The comparison of the QDM hole states and the Bi band when strain is ignored is shown in Fig.~\ref{fig:spacing-7-Bi_bands-unrelaxed_bi}. 

\paragraph*{}
Interestingly, when considering only the orbital effects of Bi, there is no change to the dot-hole-state energies with respect to Bi concentration.  This is the case for both a full GaBiAs barrier and the layered barrier with $ 4a $ of GaBiAs.  These electronic changes brought about by the Bi orbitals are confined to the alloyed region.  In contrast, changes brought about by strain are global, affecting the QDs as well as the Bi states.

\paragraph*{}
When both induced strain and orbital effects are included (see Fig.~\ref{fig:spacing-7-Bi_bands}), the barrier GaBiAs VBE is pushed much closer to the dot states than when either the induced strain alone or orbital effect alone is considered.  Both the induced strain and orbital effects push the barrier VBE toward the dot states, but it is only when the combined effect of both is considered that the barrier VBE comes close enough to the QD states for level repulsion to be significant.  For small Bi concentrations, $x$, the effect of induced strain dominates and explains the initial increase of QDM hole state energies.  For larger $x$, the combination of alloy-induced strain and Bi orbital effects are needed to explain the reversal of the QDM hole energy shifts under level repulsion.

\paragraph*{}
Although the electronic effects of Bi orbitals alone do not directly affect the QDM hole state energies, much more pronounced changes can be seen in the tunneling behavior between the two dots when the hole states of the two QDs are brought into resonance by a vertical field.  The Bi orbitals manipulate the dot wave functions to form tunneling pathways between the two dots and greatly perturb the tunnel coupling and spin-mixing in the dots.  We shall explore such behavior in the follow-up paper.


\section{Conclusion} \label{sec:conclusion}

\paragraph*{}
Hole spins in self-assembled semiconductor QDs offer intriguing possibilities as potential qubit architectures.  Hole spins in QD molecules can be manipulated by indirect optical transitions to provide all-optical control of spin initialization, manipulation, and readout. Spin initialization and manipulation requires spin mixing of holes when tunnel coupled across the dots in the QDM.  In this paper and in a follow-up paper, we study how the GaAs interdot barrier can be engineered by changing the barrier composition to enhance the spin-mixing.  Here, we investigate how to treat \ce{GaBi_xAs_{1-x}} alloys to correctly model their properties in InAs QDMs, to wit, with an atomistic TB model.  Such an atomistic treatment is necessary to describe the alloying effects of Bi, and allows for the modeling of alloy fluctuations and clustering effects.  We also showed that a VCA for the alloy fails to describe the alloy adequately.  With the atomistic TB model, we find that BAC plays a significant role in determining the valence band energies of GaBiAs.  As a consequence, band repulsion effects due to local Bi states in the barrier can modify states localized in the QDM, especially when the Bi concentration approaches 10\%.  Electron states are not strongly dependent on the presence of Bi in the barrier nor to fluctuations in the alloy configuration.  However, hole states are much more sensitive to the presence of Bi and to configurations of the alloy.  Significant reductions in the hole energies occur as the Bi concentration increases in the barrier.  This occurs both because the barrier height is reduced and because, for Bi concentrations near 10\%, the band of Bi barrier valence states pushes into the QD hole states.  We separately investigate the effects of the strain induced when the larger Bi atom replaces the smaller As atom and the effects when the As orbital energies are replaced by Bi orbital energies (i.e.zz\ the scattering induced when Bi replaces As). The induced strain is a global effect because the strain propagates from the barrier into the QDs, while the scattering off the difference between Bi and As orbitals is a more local effect.  We initially see an increase in QDM hole energies caused by the alloy-induced strain.  Then, at higher Bi concentrations, the combined effect of alloy-induced strain and Bi orbitals results in a reversal of QDM hole energy shifts, characterized by an encroaching barrier VBE leading to level repulsion between the barrier VBE and the QDM hole states.  This work has established a basis for future understanding of the spin-mixing and enhanced tunnel coupling that occurs when GaBiAs interdot barriers are used in InAs QDMs. This study shall be followed up by experimental results from our collaborators.  Then, in a subsequent paper we will show how these barriers can be used to enhance the operation of hole spin qubits in InAs QDMs.


\begin{acknowledgments}
Computational resources for this work are provided by the National Institute of Standards and Technology (NIST).  This project is supported under National Science Foundation (NSF) Grant No. DMR-1505628.
\end{acknowledgments}


\appendix


\section{Computational details for periodic and clustered alloy band-structure calculations} \label{appx:gabias}

\paragraph*{}
The band structure for \ce{GaBi_{0.25}As_{0.75}} is calculated by replacing the center As in an eight-atom unit cell of GaAs with Bi.  All the TB parameters associated with the center As atom and its bonds are replaced with Bi and GaBi, respectively.  However, all bond lengths are still taken to be that of GaAs, as the structure is not strain relaxed.  Additionally, since the lattice is unrelaxed, the TB parameters are not scaled with changes from natural bond length, despite the GaBi bonds being compressed to match GaAs.

\paragraph*{}
Finally, for Bi cluster calculations, we create square structures that are $ m $ unit cells in length, where $ m = {1, 2, 3, 4, 5} $.  The total number of atoms in our structure is $ 8 \times m^3 $, or 8, 64, 216, 512, and 1000 atoms.  Periodic conditions are identical to the case of the unit cell, but scaled to the size of the structure.  For example, the structure with length $ m = 2a $ wraps around on itself every two lattice constants in each crystal direction.  The Bi replacement begins with the anion at the center of the structure, regardless of structure size.  Up to three additional next nearest neighbor anions are replaced with Bi, in the order of $ \half\half0 $, $ \half 0 \half $, and $ 0 \half\half $.  The lattices are unrelaxed, and all the bond lengths of GaAs are kept.  TB parameters are not scaled with bond length.  Only the energy at the zone center is shown.  With this method, there are two variables affecting Bi concentration.  One is the number of Bi atoms; the other is the size of the structure.  Increasing the structure size lowers the Bi concentration directly, without increasing cluster size.  Adding Bi to the system increases the concentration, but results in a greater VBE shift than reducing the structure size by the corresponding amount.


\section{Geometry of alloy-induced strain in QDMs} \label{appx:strain_profile}

\begin{figure}[ht]
	\centering
	\subfloat[Full Barrier, ground state]{
		\includegraphics[width=0.24\textwidth]{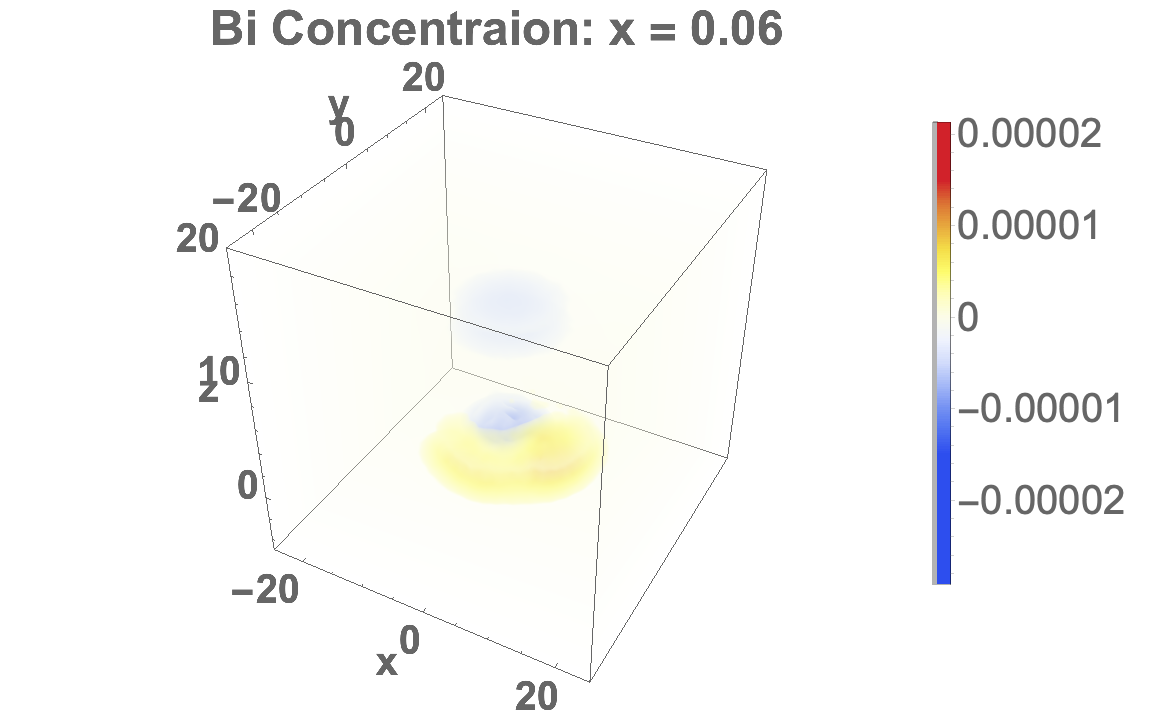}
		\label{fig-sub:wf_diff-full-ground}
	}
	\subfloat[Full Barrier, first excited state]{
		\includegraphics[width=0.24\textwidth]{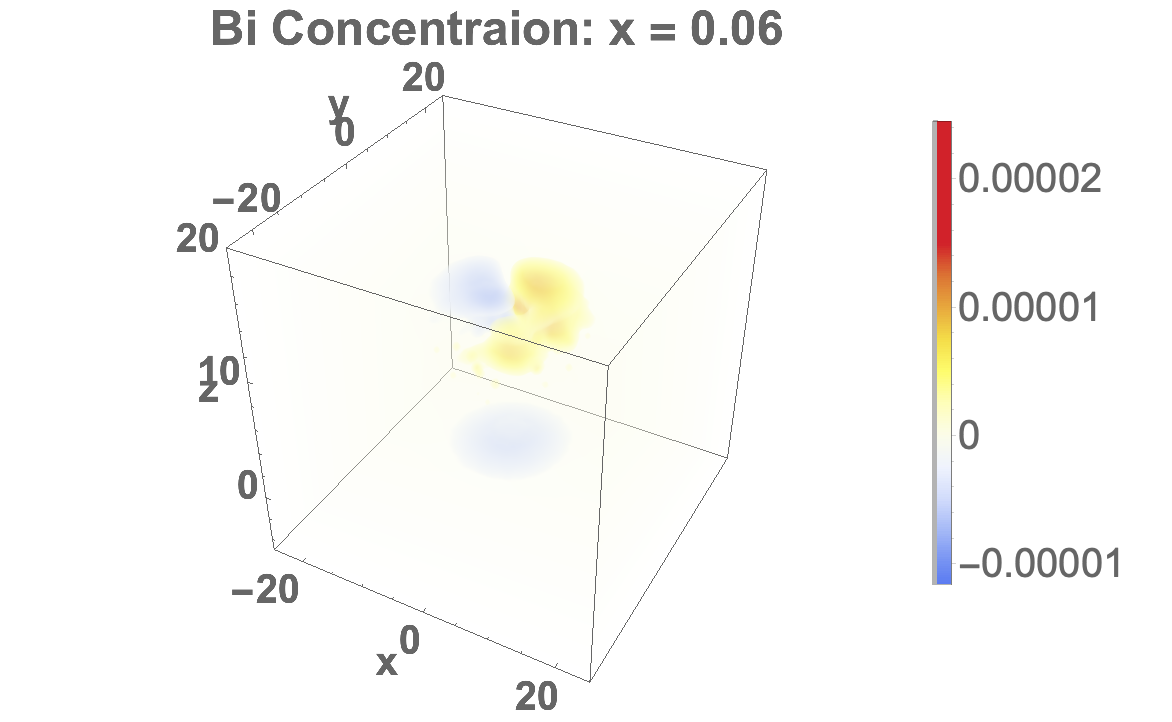}
		\label{fig-sub:wf_diff-full-excited}
	}
	\hspace{0em}
	\subfloat[Layered Barrier, ground state]{
		\includegraphics[width=0.24\textwidth]{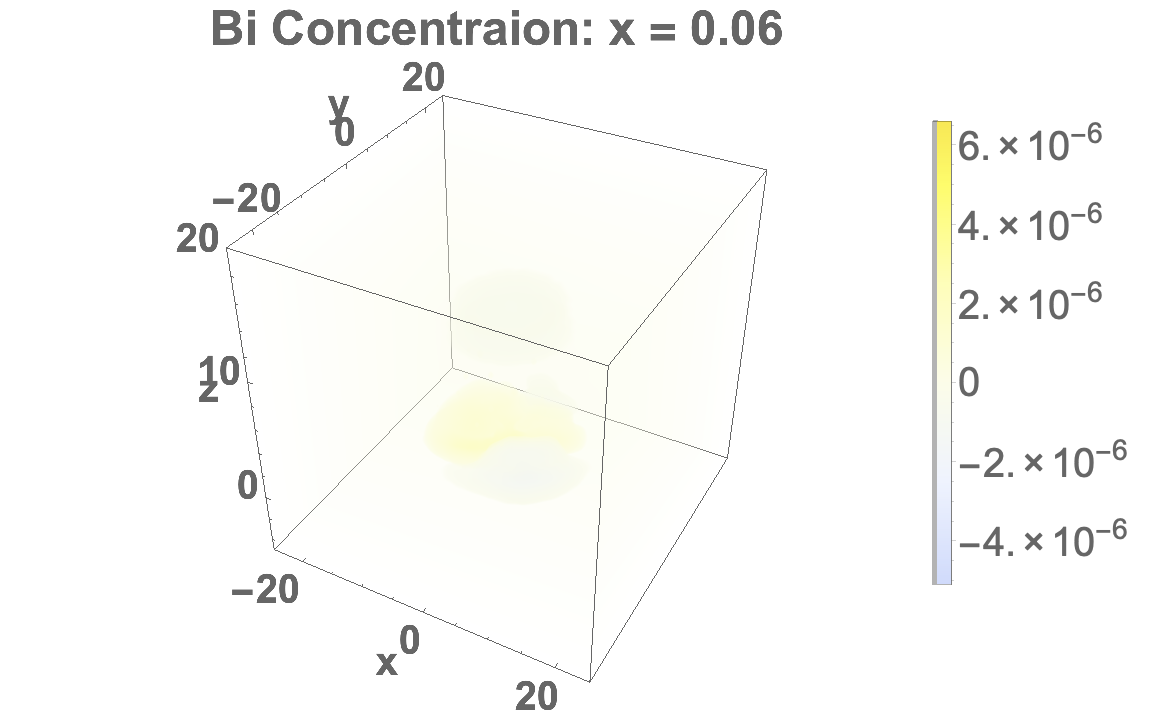}
		\label{fig-sub:wf_diff-part-ground}
	}
	\subfloat[Layered Barrier, first excited state]{
		\includegraphics[width=0.24\textwidth]{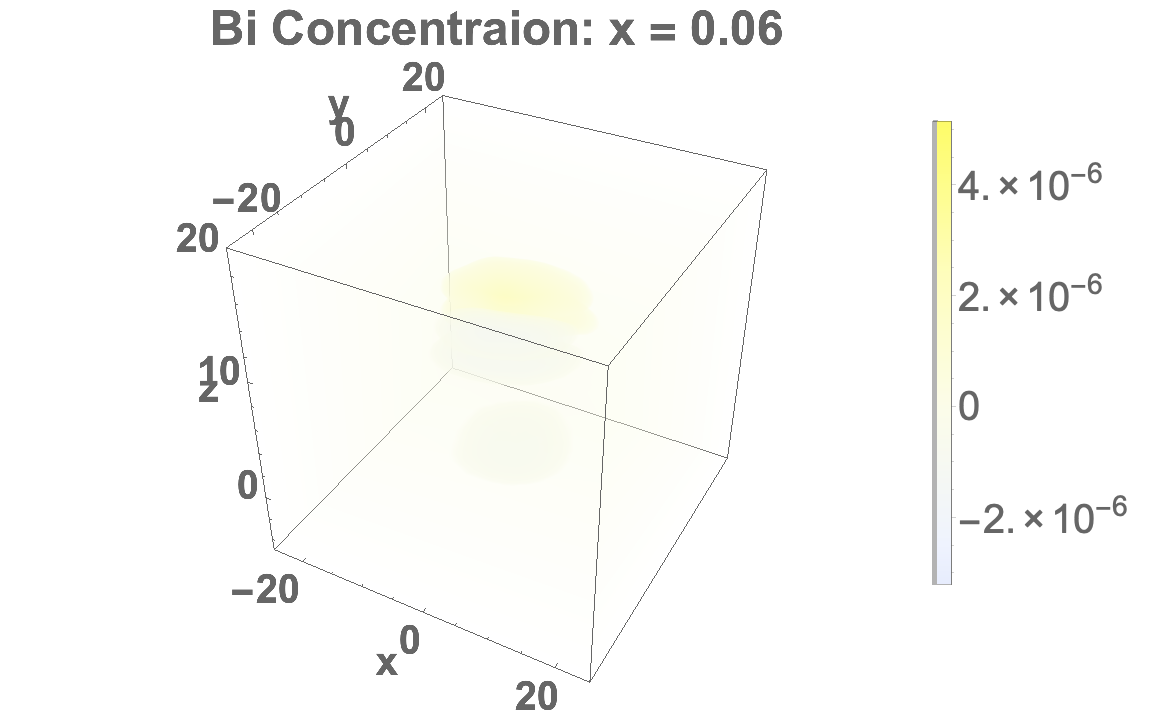}
		\label{fig-sub:wf_diff-part-excited}
	}
	\caption[Increased wave function confinement from alloy strain.]{\small (Color online)  Change in QDM hole probability when the alloy strain of a 6\% \ce{GaBi_xAs_{1-x}} alloy is applied, for (a) the ground/bottom-dot state and (b) first-excited/top-dot state of a QDM with the entire interdot region as GaBiAs, and for (c) the ground/bottom-dot state and (d) first-excited/top-dot state of a QDM with a $ 4a $ layer of GaBiAs in the interdot region.  Axes labels indicate position in lattice constants; scale bar indicates difference in wave function probability before and after strain is included.}
	\label{fig:wf_diff-strain}
\end{figure}

\paragraph*{}
Figure~\ref{fig:wf_diff-strain} shows the change in InAs QDM hole state probability after alloy strain is introduced (but with orbital differences between Bi and As ignored).  The change is calculated as the difference between the 6\% alloy case and the 0\% alloy case in Fig.~\ref{fig:spacing-7-Bi_bands-not_bi}.  The strain from the alloy confines the wave function to a single dot (bottom dot for the ground state; top dot for the first excited state), and is the reason we see an increase in QDM hole state energy at low concentrations of Bi, for both the full and layered barrier.

\begin{figure}[ht]
	\centering
	\subfloat[Full GaBiAs Barrier]{
		\includegraphics[width=0.5\textwidth]{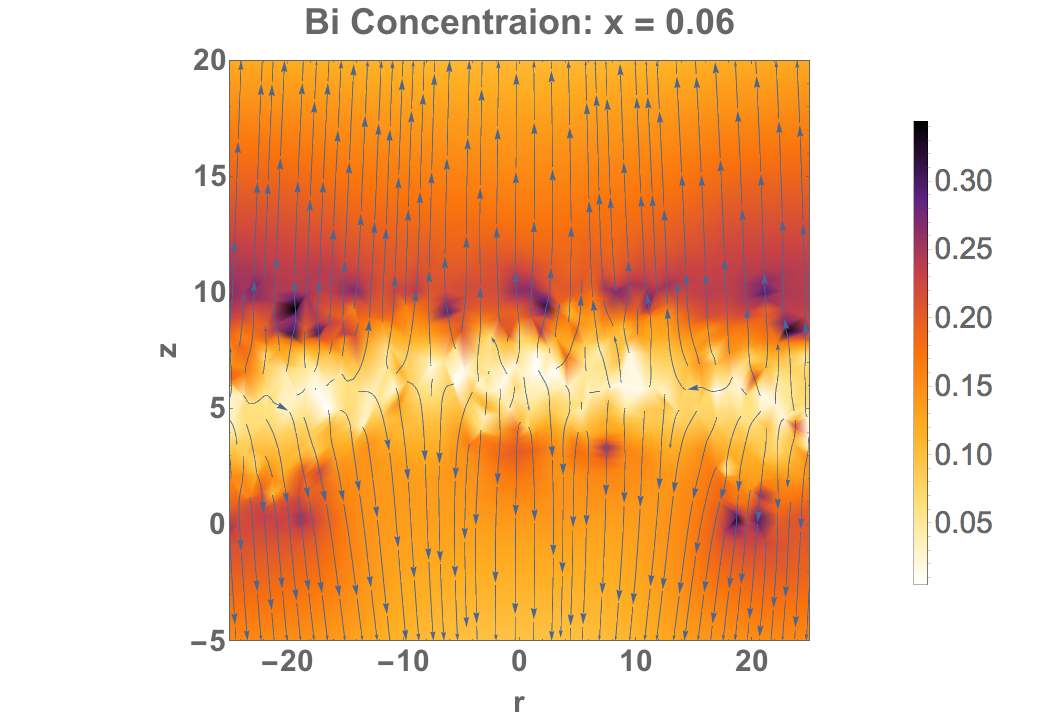}
		\label{fig-sub:strain_profile-full}
	}
	\hspace{0em}
	\subfloat[Partial GaBiAs Barrier]{
		\includegraphics[width=0.5\textwidth]{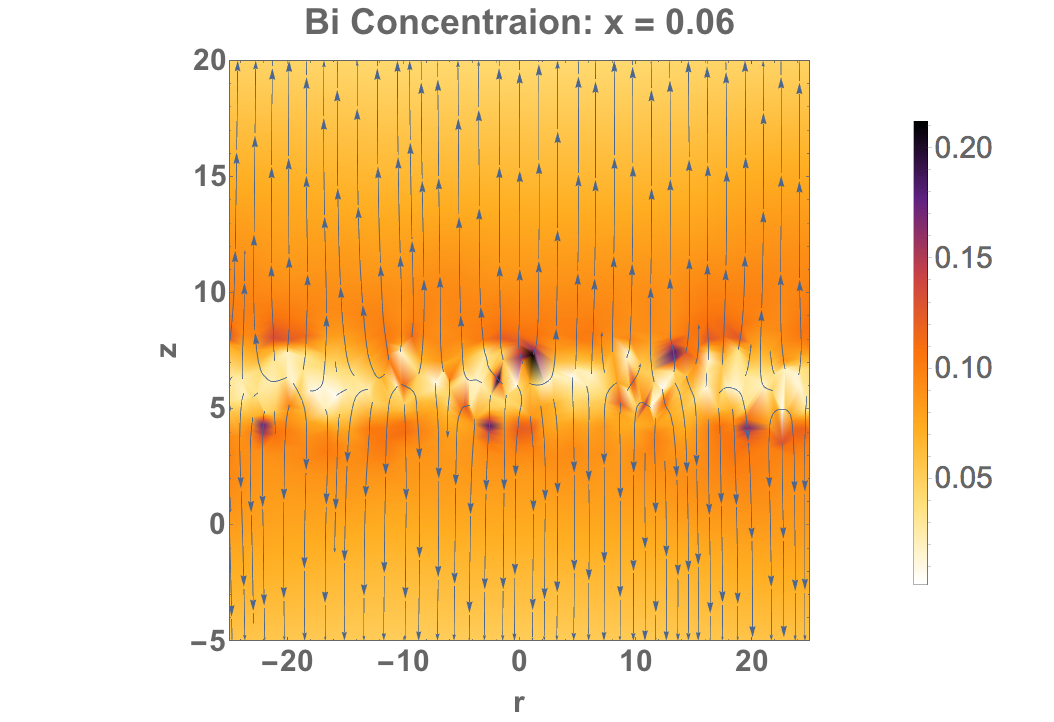}
		\label{fig-sub:strain_profile-part}
	}
	\caption[Alloy-induced strain profile.]{\small (Color online)  Shift in atomic position resulting from alloy strain for an InAs/GaBiAs QDM  with (a) the entire interdot region as \ce{GaBi_{0.06}As_{0.94}} or (b) the interdot region containing a $ 4a $ layer of \ce{GaBi_{0.06}As_{0.94}}.  Arrows represent the direction of position shift.  Backdrop represents magnitude of shift; scale bar in units of \AA.  2D slice taken along $ 110 $ plane (distance in lattice constants $a$).}
	\label{fig:strain_profile}
\end{figure}

\paragraph*{}
Figure~\ref{fig-sub:strain_profile-full} shows the strain from the alloy region clearly propagating away from the alloy.  When the entire interdot region is alloyed, the Bi surrounds the bottom dot, and the strain is concentrated outside of the bottom dot near the sidewalls.  The top dot, on the other hand, has the strain accumulating uniformly beneath the wetting layer and propagating through the dot.  Therefore, the top dot receives much more of the alloy-induced strain, whereas the bottom dot has strain surrounding it but not penetrating the dot itself.  The result is increased difference between the bottom (ground) and top (first excited) state energies seen in Fig.~\ref{fig:spacing-7-Bi_bands-not_bi-thickness}.

\paragraph*{}
For a symmetrically layered interdot barrier, with a $ 4a $ layer of GaBiAs sandwiched between GaAs, the strain is able to propagate above and below the alloy region equally [see Fig.~\ref{fig-sub:strain_profile-part}].  The alloy-induced strain is able to affect both dots in a fashion similar to how the top dot was affected in the full barrier case.  However, due to the distance between the alloy layer and the dots, the strain is diminished by the time it reaches the dot, resulting in less of an energy shift in comparison with the top dot in the full barrier case in Fig.~\ref{fig:spacing-7-Bi_bands-not_bi-thickness}.

\twocolumngrid


\bibliography{bibliography}

\begin{thebibliography}{46}%
\makeatletter
\providecommand \@ifxundefined [1]{%
 \@ifx{#1\undefined}
}%
\providecommand \@ifnum [1]{%
 \ifnum #1\expandafter \@firstoftwo
 \else \expandafter \@secondoftwo
 \fi
}%
\providecommand \@ifx [1]{%
 \ifx #1\expandafter \@firstoftwo
 \else \expandafter \@secondoftwo
 \fi
}%
\providecommand \natexlab [1]{#1}%
\providecommand \enquote  [1]{``#1''}%
\providecommand \bibnamefont  [1]{#1}%
\providecommand \bibfnamefont [1]{#1}%
\providecommand \citenamefont [1]{#1}%
\providecommand \href@noop [0]{\@secondoftwo}%
\providecommand \href [0]{\begingroup \@sanitize@url \@href}%
\providecommand \@href[1]{\@@startlink{#1}\@@href}%
\providecommand \@@href[1]{\endgroup#1\@@endlink}%
\providecommand \@sanitize@url [0]{\catcode `\\12\catcode `\$12\catcode
  `\&12\catcode `\#12\catcode `\^12\catcode `\_12\catcode `\%12\relax}%
\providecommand \@@startlink[1]{}%
\providecommand \@@endlink[0]{}%
\providecommand \url  [0]{\begingroup\@sanitize@url \@url }%
\providecommand \@url [1]{\endgroup\@href {#1}{\urlprefix }}%
\providecommand \urlprefix  [0]{URL }%
\providecommand \Eprint [0]{\href }%
\providecommand \doibase [0]{http://dx.doi.org/}%
\providecommand \selectlanguage [0]{\@gobble}%
\providecommand \bibinfo  [0]{\@secondoftwo}%
\providecommand \bibfield  [0]{\@secondoftwo}%
\providecommand \translation [1]{[#1]}%
\providecommand \BibitemOpen [0]{}%
\providecommand \bibitemStop [0]{}%
\providecommand \bibitemNoStop [0]{.\EOS\space}%
\providecommand \EOS [0]{\spacefactor3000\relax}%
\providecommand \BibitemShut  [1]{\csname bibitem#1\endcsname}%
\let\auto@bib@innerbib\@empty
\bibitem [{\citenamefont {Ladd}\ \emph {et~al.}(2010)\citenamefont {Ladd},
  \citenamefont {Jelezko}, \citenamefont {Laflamme}, \citenamefont {Nakamura},
  \citenamefont {Monroe},\ and\ \citenamefont {O{
  extquoteright}Brien}}]{ladd_quantum_2010}%
  \BibitemOpen
  \bibfield  {author} {\bibinfo {author} {\bibfnamefont {T.~D.}\ \bibnamefont
  {Ladd}}, \bibinfo {author} {\bibfnamefont {F.}~\bibnamefont {Jelezko}},
  \bibinfo {author} {\bibfnamefont {R.}~\bibnamefont {Laflamme}}, \bibinfo
  {author} {\bibfnamefont {Y.}~\bibnamefont {Nakamura}}, \bibinfo {author}
  {\bibfnamefont {C.}~\bibnamefont {Monroe}}, \ and\ \bibinfo {author}
  {\bibfnamefont {J.~L.}\ \bibnamefont {O{ extquoteright}Brien}},\ }\href
  {\doibase 10.1038/nature08812} {\bibfield  {journal} {\bibinfo  {journal}
  {Nature}\ }\textbf {\bibinfo {volume} {464}},\ \bibinfo {pages} {45}
  (\bibinfo {year} {2010})}\BibitemShut {NoStop}%
\bibitem [{\citenamefont {Morton}\ \emph {et~al.}(2011)\citenamefont {Morton},
  \citenamefont {McCamey}, \citenamefont {Eriksson},\ and\ \citenamefont
  {Lyon}}]{morton_embracing_2011}%
  \BibitemOpen
  \bibfield  {author} {\bibinfo {author} {\bibfnamefont {J.~J.~L.}\
  \bibnamefont {Morton}}, \bibinfo {author} {\bibfnamefont {D.~R.}\
  \bibnamefont {McCamey}}, \bibinfo {author} {\bibfnamefont {M.~A.}\
  \bibnamefont {Eriksson}}, \ and\ \bibinfo {author} {\bibfnamefont {S.~A.}\
  \bibnamefont {Lyon}},\ }\href {\doibase 10.1038/nature10681} {\bibfield
  {journal} {\bibinfo  {journal} {Nature}\ }\textbf {\bibinfo {volume} {479}},\
  \bibinfo {pages} {345} (\bibinfo {year} {2011})}\BibitemShut {NoStop}%
\bibitem [{\citenamefont {Awschalom}\ \emph {et~al.}(2013)\citenamefont
  {Awschalom}, \citenamefont {Bassett}, \citenamefont {Dzurak}, \citenamefont
  {Hu},\ and\ \citenamefont {Petta}}]{awschalom_quantum_2013}%
  \BibitemOpen
  \bibfield  {author} {\bibinfo {author} {\bibfnamefont {D.~D.}\ \bibnamefont
  {Awschalom}}, \bibinfo {author} {\bibfnamefont {L.~C.}\ \bibnamefont
  {Bassett}}, \bibinfo {author} {\bibfnamefont {A.~S.}\ \bibnamefont {Dzurak}},
  \bibinfo {author} {\bibfnamefont {E.~L.}\ \bibnamefont {Hu}}, \ and\ \bibinfo
  {author} {\bibfnamefont {J.~R.}\ \bibnamefont {Petta}},\ }\href {\doibase
  10.1126/science.1231364} {\bibfield  {journal} {\bibinfo  {journal}
  {Science}\ }\textbf {\bibinfo {volume} {339}},\ \bibinfo {pages} {1174}
  (\bibinfo {year} {2013})}\BibitemShut {NoStop}%
\bibitem [{\citenamefont {Bonadeo}\ \emph {et~al.}(1998)\citenamefont
  {Bonadeo}, \citenamefont {Erland}, \citenamefont {Gammon}, \citenamefont
  {Park}, \citenamefont {Katzer},\ and\ \citenamefont
  {Steel}}]{bonadeo_coherent_1998}%
  \BibitemOpen
  \bibfield  {author} {\bibinfo {author} {\bibfnamefont {N.~H.}\ \bibnamefont
  {Bonadeo}}, \bibinfo {author} {\bibfnamefont {J.}~\bibnamefont {Erland}},
  \bibinfo {author} {\bibfnamefont {D.}~\bibnamefont {Gammon}}, \bibinfo
  {author} {\bibfnamefont {D.}~\bibnamefont {Park}}, \bibinfo {author}
  {\bibfnamefont {D.~S.}\ \bibnamefont {Katzer}}, \ and\ \bibinfo {author}
  {\bibfnamefont {D.~G.}\ \bibnamefont {Steel}},\ }\href {\doibase
  10.1126/science.282.5393.1473} {\bibfield  {journal} {\bibinfo  {journal}
  {Science}\ }\textbf {\bibinfo {volume} {282}},\ \bibinfo {pages} {1473}
  (\bibinfo {year} {1998})}\BibitemShut {NoStop}%
\bibitem [{\citenamefont {Merkulov}\ \emph {et~al.}(2002)\citenamefont
  {Merkulov}, \citenamefont {Efros},\ and\ \citenamefont
  {Rosen}}]{merkulov_electron_2002}%
  \BibitemOpen
  \bibfield  {author} {\bibinfo {author} {\bibfnamefont {I.~A.}\ \bibnamefont
  {Merkulov}}, \bibinfo {author} {\bibfnamefont {A.~L.}\ \bibnamefont {Efros}},
  \ and\ \bibinfo {author} {\bibfnamefont {M.}~\bibnamefont {Rosen}},\ }\href
  {\doibase 10.1103/PhysRevB.65.205309} {\bibfield  {journal} {\bibinfo
  {journal} {Phys. Rev. B}\ }\textbf {\bibinfo {volume} {65}},\ \bibinfo
  {pages} {205309} (\bibinfo {year} {2002})}\BibitemShut {NoStop}%
\bibitem [{\citenamefont {Khaetskii}\ and\ \citenamefont
  {Nazarov}(2000)}]{khaetskii_spin_2000}%
  \BibitemOpen
  \bibfield  {author} {\bibinfo {author} {\bibfnamefont {A.~V.}\ \bibnamefont
  {Khaetskii}}\ and\ \bibinfo {author} {\bibfnamefont {Y.~V.}\ \bibnamefont
  {Nazarov}},\ }\href {\doibase 10.1103/PhysRevB.61.12639} {\bibfield
  {journal} {\bibinfo  {journal} {Phys. Rev. B}\ }\textbf {\bibinfo {volume}
  {61}},\ \bibinfo {pages} {12639} (\bibinfo {year} {2000})}\BibitemShut
  {NoStop}%
\bibitem [{\citenamefont {Khaetskii}\ \emph {et~al.}(2002)\citenamefont
  {Khaetskii}, \citenamefont {Loss},\ and\ \citenamefont
  {Glazman}}]{khaetskii_electron_2002}%
  \BibitemOpen
  \bibfield  {author} {\bibinfo {author} {\bibfnamefont {A.~V.}\ \bibnamefont
  {Khaetskii}}, \bibinfo {author} {\bibfnamefont {D.}~\bibnamefont {Loss}}, \
  and\ \bibinfo {author} {\bibfnamefont {L.}~\bibnamefont {Glazman}},\ }\href
  {\doibase 10.1103/PhysRevLett.88.186802} {\bibfield  {journal} {\bibinfo
  {journal} {Phys. Rev. Lett.}\ }\textbf {\bibinfo {volume} {88}},\ \bibinfo
  {pages} {186802} (\bibinfo {year} {2002})}\BibitemShut {NoStop}%
\bibitem [{\citenamefont {Brunner}\ \emph {et~al.}(2009)\citenamefont
  {Brunner}, \citenamefont {Gerardot}, \citenamefont {Dalgarno}, \citenamefont
  {W{\"u}st}, \citenamefont {Karrai}, \citenamefont {Stoltz}, \citenamefont
  {Petroff},\ and\ \citenamefont {Warburton}}]{brunner_coherent_2009}%
  \BibitemOpen
  \bibfield  {author} {\bibinfo {author} {\bibfnamefont {D.}~\bibnamefont
  {Brunner}}, \bibinfo {author} {\bibfnamefont {B.~D.}\ \bibnamefont
  {Gerardot}}, \bibinfo {author} {\bibfnamefont {P.~A.}\ \bibnamefont
  {Dalgarno}}, \bibinfo {author} {\bibfnamefont {G.}~\bibnamefont {W{\"u}st}},
  \bibinfo {author} {\bibfnamefont {K.}~\bibnamefont {Karrai}}, \bibinfo
  {author} {\bibfnamefont {N.~G.}\ \bibnamefont {Stoltz}}, \bibinfo {author}
  {\bibfnamefont {P.~M.}\ \bibnamefont {Petroff}}, \ and\ \bibinfo {author}
  {\bibfnamefont {R.~J.}\ \bibnamefont {Warburton}},\ }\href {\doibase
  10.1126/science.1173684} {\bibfield  {journal} {\bibinfo  {journal}
  {Science}\ }\textbf {\bibinfo {volume} {325}},\ \bibinfo {pages} {70}
  (\bibinfo {year} {2009})}\BibitemShut {NoStop}%
\bibitem [{\citenamefont {Eble}\ \emph {et~al.}(2009)\citenamefont {Eble},
  \citenamefont {Testelin}, \citenamefont {Desfonds}, \citenamefont
  {Bernardot}, \citenamefont {Balocchi}, \citenamefont {Amand}, \citenamefont
  {Miard}, \citenamefont {Lema{\^i}tre}, \citenamefont {Marie},\ and\
  \citenamefont {Chamarro}}]{eble_holenuclear_2009}%
  \BibitemOpen
  \bibfield  {author} {\bibinfo {author} {\bibfnamefont {B.}~\bibnamefont
  {Eble}}, \bibinfo {author} {\bibfnamefont {C.}~\bibnamefont {Testelin}},
  \bibinfo {author} {\bibfnamefont {P.}~\bibnamefont {Desfonds}}, \bibinfo
  {author} {\bibfnamefont {F.}~\bibnamefont {Bernardot}}, \bibinfo {author}
  {\bibfnamefont {A.}~\bibnamefont {Balocchi}}, \bibinfo {author}
  {\bibfnamefont {T.}~\bibnamefont {Amand}}, \bibinfo {author} {\bibfnamefont
  {A.}~\bibnamefont {Miard}}, \bibinfo {author} {\bibfnamefont
  {A.}~\bibnamefont {Lema{\^i}tre}}, \bibinfo {author} {\bibfnamefont
  {X.}~\bibnamefont {Marie}}, \ and\ \bibinfo {author} {\bibfnamefont
  {M.}~\bibnamefont {Chamarro}},\ }\href {\doibase
  10.1103/PhysRevLett.102.146601} {\bibfield  {journal} {\bibinfo  {journal}
  {Physical Review Letters}\ }\textbf {\bibinfo {volume} {102}},\ \bibinfo
  {pages} {146601} (\bibinfo {year} {2009})}\BibitemShut {NoStop}%
\bibitem [{\citenamefont {Testelin}\ \emph {et~al.}(2009)\citenamefont
  {Testelin}, \citenamefont {Bernardot}, \citenamefont {Eble},\ and\
  \citenamefont {Chamarro}}]{testelin_hole-spin_2009}%
  \BibitemOpen
  \bibfield  {author} {\bibinfo {author} {\bibfnamefont {C.}~\bibnamefont
  {Testelin}}, \bibinfo {author} {\bibfnamefont {F.}~\bibnamefont {Bernardot}},
  \bibinfo {author} {\bibfnamefont {B.}~\bibnamefont {Eble}}, \ and\ \bibinfo
  {author} {\bibfnamefont {M.}~\bibnamefont {Chamarro}},\ }\href {\doibase
  10.1103/PhysRevB.79.195440} {\bibfield  {journal} {\bibinfo  {journal} {Phys.
  Rev. B}\ }\textbf {\bibinfo {volume} {79}},\ \bibinfo {pages} {195440}
  (\bibinfo {year} {2009})}\BibitemShut {NoStop}%
\bibitem [{\citenamefont {Economou}\ \emph {et~al.}(2012)\citenamefont
  {Economou}, \citenamefont {Climente}, \citenamefont {Badolato}, \citenamefont
  {Bracker}, \citenamefont {Gammon},\ and\ \citenamefont
  {Doty}}]{economou_scalable_2012}%
  \BibitemOpen
  \bibfield  {author} {\bibinfo {author} {\bibfnamefont {S.~E.}\ \bibnamefont
  {Economou}}, \bibinfo {author} {\bibfnamefont {J.~I.}\ \bibnamefont
  {Climente}}, \bibinfo {author} {\bibfnamefont {A.}~\bibnamefont {Badolato}},
  \bibinfo {author} {\bibfnamefont {A.~S.}\ \bibnamefont {Bracker}}, \bibinfo
  {author} {\bibfnamefont {D.}~\bibnamefont {Gammon}}, \ and\ \bibinfo {author}
  {\bibfnamefont {M.~F.}\ \bibnamefont {Doty}},\ }\href {\doibase
  10.1103/PhysRevB.86.085319} {\bibfield  {journal} {\bibinfo  {journal} {Phys.
  Rev. B}\ }\textbf {\bibinfo {volume} {86}},\ \bibinfo {pages} {085319}
  (\bibinfo {year} {2012})}\BibitemShut {NoStop}%
\bibitem [{\citenamefont {Bayer}\ \emph {et~al.}(2001)\citenamefont {Bayer},
  \citenamefont {Hawrylak}, \citenamefont {Hinzer}, \citenamefont {Fafard},
  \citenamefont {Korkusinski}, \citenamefont {Wasilewski}, \citenamefont
  {Stern},\ and\ \citenamefont {Forchel}}]{bayer_coupling_2001}%
  \BibitemOpen
  \bibfield  {author} {\bibinfo {author} {\bibfnamefont {M.}~\bibnamefont
  {Bayer}}, \bibinfo {author} {\bibfnamefont {P.}~\bibnamefont {Hawrylak}},
  \bibinfo {author} {\bibfnamefont {K.}~\bibnamefont {Hinzer}}, \bibinfo
  {author} {\bibfnamefont {S.}~\bibnamefont {Fafard}}, \bibinfo {author}
  {\bibfnamefont {M.}~\bibnamefont {Korkusinski}}, \bibinfo {author}
  {\bibfnamefont {Z.~R.}\ \bibnamefont {Wasilewski}}, \bibinfo {author}
  {\bibfnamefont {O.}~\bibnamefont {Stern}}, \ and\ \bibinfo {author}
  {\bibfnamefont {A.}~\bibnamefont {Forchel}},\ }\href {\doibase
  10.1126/science.291.5503.451} {\bibfield  {journal} {\bibinfo  {journal}
  {Science}\ }\textbf {\bibinfo {volume} {291}},\ \bibinfo {pages} {451}
  (\bibinfo {year} {2001})}\BibitemShut {NoStop}%
\bibitem [{\citenamefont {Krenner}\ \emph {et~al.}(2005)\citenamefont
  {Krenner}, \citenamefont {Sabathil}, \citenamefont {Clark}, \citenamefont
  {Kress}, \citenamefont {Schuh}, \citenamefont {Bichler}, \citenamefont
  {Abstreiter},\ and\ \citenamefont {Finley}}]{krenner_direct_2005}%
  \BibitemOpen
  \bibfield  {author} {\bibinfo {author} {\bibfnamefont {H.~J.}\ \bibnamefont
  {Krenner}}, \bibinfo {author} {\bibfnamefont {M.}~\bibnamefont {Sabathil}},
  \bibinfo {author} {\bibfnamefont {E.~C.}\ \bibnamefont {Clark}}, \bibinfo
  {author} {\bibfnamefont {A.}~\bibnamefont {Kress}}, \bibinfo {author}
  {\bibfnamefont {D.}~\bibnamefont {Schuh}}, \bibinfo {author} {\bibfnamefont
  {M.}~\bibnamefont {Bichler}}, \bibinfo {author} {\bibfnamefont
  {G.}~\bibnamefont {Abstreiter}}, \ and\ \bibinfo {author} {\bibfnamefont
  {J.~J.}\ \bibnamefont {Finley}},\ }\href {\doibase
  10.1103/PhysRevLett.94.057402} {\bibfield  {journal} {\bibinfo  {journal}
  {Phys. Rev. Lett.}\ }\textbf {\bibinfo {volume} {94}},\ \bibinfo {pages}
  {057402} (\bibinfo {year} {2005})}\BibitemShut {NoStop}%
\bibitem [{\citenamefont {Ortner}\ \emph {et~al.}(2005)\citenamefont {Ortner},
  \citenamefont {Bayer}, \citenamefont {Lyanda-Geller}, \citenamefont
  {Reinecke}, \citenamefont {Kress}, \citenamefont {Reithmaier},\ and\
  \citenamefont {Forchel}}]{ortner_control_2005}%
  \BibitemOpen
  \bibfield  {author} {\bibinfo {author} {\bibfnamefont {G.}~\bibnamefont
  {Ortner}}, \bibinfo {author} {\bibfnamefont {M.}~\bibnamefont {Bayer}},
  \bibinfo {author} {\bibfnamefont {Y.}~\bibnamefont {Lyanda-Geller}}, \bibinfo
  {author} {\bibfnamefont {T.~L.}\ \bibnamefont {Reinecke}}, \bibinfo {author}
  {\bibfnamefont {A.}~\bibnamefont {Kress}}, \bibinfo {author} {\bibfnamefont
  {J.~P.}\ \bibnamefont {Reithmaier}}, \ and\ \bibinfo {author} {\bibfnamefont
  {A.}~\bibnamefont {Forchel}},\ }\href {\doibase
  10.1103/PhysRevLett.94.157401} {\bibfield  {journal} {\bibinfo  {journal}
  {Phys. Rev. Lett.}\ }\textbf {\bibinfo {volume} {94}},\ \bibinfo {pages}
  {157401} (\bibinfo {year} {2005})}\BibitemShut {NoStop}%
\bibitem [{\citenamefont {Stinaff}\ \emph {et~al.}(2006)\citenamefont
  {Stinaff}, \citenamefont {Scheibner}, \citenamefont {Bracker}, \citenamefont
  {Ponomarev}, \citenamefont {Korenev}, \citenamefont {Ware}, \citenamefont
  {Doty}, \citenamefont {Reinecke},\ and\ \citenamefont
  {Gammon}}]{stinaff_optical_2006}%
  \BibitemOpen
  \bibfield  {author} {\bibinfo {author} {\bibfnamefont {E.~A.}\ \bibnamefont
  {Stinaff}}, \bibinfo {author} {\bibfnamefont {M.}~\bibnamefont {Scheibner}},
  \bibinfo {author} {\bibfnamefont {A.~S.}\ \bibnamefont {Bracker}}, \bibinfo
  {author} {\bibfnamefont {I.~V.}\ \bibnamefont {Ponomarev}}, \bibinfo {author}
  {\bibfnamefont {V.~L.}\ \bibnamefont {Korenev}}, \bibinfo {author}
  {\bibfnamefont {M.~E.}\ \bibnamefont {Ware}}, \bibinfo {author}
  {\bibfnamefont {M.~F.}\ \bibnamefont {Doty}}, \bibinfo {author}
  {\bibfnamefont {T.~L.}\ \bibnamefont {Reinecke}}, \ and\ \bibinfo {author}
  {\bibfnamefont {D.}~\bibnamefont {Gammon}},\ }\href {\doibase
  10.1126/science.1121189} {\bibfield  {journal} {\bibinfo  {journal}
  {Science}\ }\textbf {\bibinfo {volume} {311}},\ \bibinfo {pages} {636}
  (\bibinfo {year} {2006})}\BibitemShut {NoStop}%
\bibitem [{\citenamefont {Doty}\ \emph {et~al.}(2008)\citenamefont {Doty},
  \citenamefont {Scheibner}, \citenamefont {Bracker}, \citenamefont
  {Ponomarev}, \citenamefont {Reinecke},\ and\ \citenamefont
  {Gammon}}]{doty_optical_2008}%
  \BibitemOpen
  \bibfield  {author} {\bibinfo {author} {\bibfnamefont {M.~F.}\ \bibnamefont
  {Doty}}, \bibinfo {author} {\bibfnamefont {M.}~\bibnamefont {Scheibner}},
  \bibinfo {author} {\bibfnamefont {A.~S.}\ \bibnamefont {Bracker}}, \bibinfo
  {author} {\bibfnamefont {I.~V.}\ \bibnamefont {Ponomarev}}, \bibinfo {author}
  {\bibfnamefont {T.~L.}\ \bibnamefont {Reinecke}}, \ and\ \bibinfo {author}
  {\bibfnamefont {D.}~\bibnamefont {Gammon}},\ }\href {\doibase
  10.1103/PhysRevB.78.115316} {\bibfield  {journal} {\bibinfo  {journal} {Phys.
  Rev. B}\ }\textbf {\bibinfo {volume} {78}},\ \bibinfo {pages} {115316}
  (\bibinfo {year} {2008})}\BibitemShut {NoStop}%
\bibitem [{\citenamefont {Doty}\ \emph {et~al.}(2009)\citenamefont {Doty},
  \citenamefont {Climente}, \citenamefont {Korkusinski}, \citenamefont
  {Scheibner}, \citenamefont {Bracker}, \citenamefont {Hawrylak},\ and\
  \citenamefont {Gammon}}]{doty_antibonding_2009}%
  \BibitemOpen
  \bibfield  {author} {\bibinfo {author} {\bibfnamefont {M.~F.}\ \bibnamefont
  {Doty}}, \bibinfo {author} {\bibfnamefont {J.~I.}\ \bibnamefont {Climente}},
  \bibinfo {author} {\bibfnamefont {M.}~\bibnamefont {Korkusinski}}, \bibinfo
  {author} {\bibfnamefont {M.}~\bibnamefont {Scheibner}}, \bibinfo {author}
  {\bibfnamefont {A.~S.}\ \bibnamefont {Bracker}}, \bibinfo {author}
  {\bibfnamefont {P.}~\bibnamefont {Hawrylak}}, \ and\ \bibinfo {author}
  {\bibfnamefont {D.}~\bibnamefont {Gammon}},\ }\href {\doibase
  10.1103/PhysRevLett.102.047401} {\bibfield  {journal} {\bibinfo  {journal}
  {Phys. Rev. Lett.}\ }\textbf {\bibinfo {volume} {102}},\ \bibinfo {pages}
  {047401} (\bibinfo {year} {2009})}\BibitemShut {NoStop}%
\bibitem [{\citenamefont {Doty}\ \emph {et~al.}(2010)\citenamefont {Doty},
  \citenamefont {Climente}, \citenamefont {Greilich}, \citenamefont {Yakes},
  \citenamefont {Bracker},\ and\ \citenamefont {Gammon}}]{doty_hole-spin_2010}%
  \BibitemOpen
  \bibfield  {author} {\bibinfo {author} {\bibfnamefont {M.~F.}\ \bibnamefont
  {Doty}}, \bibinfo {author} {\bibfnamefont {J.~I.}\ \bibnamefont {Climente}},
  \bibinfo {author} {\bibfnamefont {A.}~\bibnamefont {Greilich}}, \bibinfo
  {author} {\bibfnamefont {M.}~\bibnamefont {Yakes}}, \bibinfo {author}
  {\bibfnamefont {A.~S.}\ \bibnamefont {Bracker}}, \ and\ \bibinfo {author}
  {\bibfnamefont {D.}~\bibnamefont {Gammon}},\ }\href {\doibase
  10.1103/PhysRevB.81.035308} {\bibfield  {journal} {\bibinfo  {journal} {Phys.
  Rev. B}\ }\textbf {\bibinfo {volume} {81}},\ \bibinfo {pages} {035308}
  (\bibinfo {year} {2010})}\BibitemShut {NoStop}%
\bibitem [{\citenamefont {Liu}\ \emph {et~al.}(2011)\citenamefont {Liu},
  \citenamefont {Sanwlani}, \citenamefont {Hazbun}, \citenamefont {Kolodzey},
  \citenamefont {Bracker}, \citenamefont {Gammon},\ and\ \citenamefont
  {Doty}}]{liu_situ_2011}%
  \BibitemOpen
  \bibfield  {author} {\bibinfo {author} {\bibfnamefont {W.}~\bibnamefont
  {Liu}}, \bibinfo {author} {\bibfnamefont {S.}~\bibnamefont {Sanwlani}},
  \bibinfo {author} {\bibfnamefont {R.}~\bibnamefont {Hazbun}}, \bibinfo
  {author} {\bibfnamefont {J.}~\bibnamefont {Kolodzey}}, \bibinfo {author}
  {\bibfnamefont {A.~S.}\ \bibnamefont {Bracker}}, \bibinfo {author}
  {\bibfnamefont {D.}~\bibnamefont {Gammon}}, \ and\ \bibinfo {author}
  {\bibfnamefont {M.~F.}\ \bibnamefont {Doty}},\ }\href {\doibase
  10.1103/PhysRevB.84.121304} {\bibfield  {journal} {\bibinfo  {journal} {Phys.
  Rev. B}\ }\textbf {\bibinfo {volume} {84}},\ \bibinfo {pages} {121304}
  (\bibinfo {year} {2011})}\BibitemShut {NoStop}%
\bibitem [{\citenamefont {Rajadell}\ \emph {et~al.}(2013)\citenamefont
  {Rajadell}, \citenamefont {Climente},\ and\ \citenamefont
  {Planelles}}]{rajadell_large_2013}%
  \BibitemOpen
  \bibfield  {author} {\bibinfo {author} {\bibfnamefont {F.}~\bibnamefont
  {Rajadell}}, \bibinfo {author} {\bibfnamefont {J.~I.}\ \bibnamefont
  {Climente}}, \ and\ \bibinfo {author} {\bibfnamefont {J.}~\bibnamefont
  {Planelles}},\ }\href {\doibase 10.1063/1.4823458} {\bibfield  {journal}
  {\bibinfo  {journal} {Appl. Phys. Lett.}\ }\textbf {\bibinfo {volume}
  {103}},\ \bibinfo {pages} {132105} (\bibinfo {year} {2013})}\BibitemShut
  {NoStop}%
\bibitem [{\citenamefont {Planelles}\ \emph {et~al.}(2015)\citenamefont
  {Planelles}, \citenamefont {Rajadell},\ and\ \citenamefont
  {Climente}}]{planelles_symmetry-induced_2015}%
  \BibitemOpen
  \bibfield  {author} {\bibinfo {author} {\bibfnamefont {J.}~\bibnamefont
  {Planelles}}, \bibinfo {author} {\bibfnamefont {F.}~\bibnamefont {Rajadell}},
  \ and\ \bibinfo {author} {\bibfnamefont {J.~I.}\ \bibnamefont {Climente}},\
  }\href {\doibase 10.1103/PhysRevB.92.041302} {\bibfield  {journal} {\bibinfo
  {journal} {Phys. Rev. B}\ }\textbf {\bibinfo {volume} {92}},\ \bibinfo
  {pages} {041302} (\bibinfo {year} {2015})}\BibitemShut {NoStop}%
\bibitem [{\citenamefont {Doty}\ \emph {et~al.}(2006)\citenamefont {Doty},
  \citenamefont {Scheibner}, \citenamefont {Ponomarev}, \citenamefont
  {Stinaff}, \citenamefont {Bracker}, \citenamefont {Korenev}, \citenamefont
  {Reinecke},\ and\ \citenamefont {Gammon}}]{doty_electrically_2006}%
  \BibitemOpen
  \bibfield  {author} {\bibinfo {author} {\bibfnamefont {M.~F.}\ \bibnamefont
  {Doty}}, \bibinfo {author} {\bibfnamefont {M.}~\bibnamefont {Scheibner}},
  \bibinfo {author} {\bibfnamefont {I.~V.}\ \bibnamefont {Ponomarev}}, \bibinfo
  {author} {\bibfnamefont {E.~A.}\ \bibnamefont {Stinaff}}, \bibinfo {author}
  {\bibfnamefont {A.~S.}\ \bibnamefont {Bracker}}, \bibinfo {author}
  {\bibfnamefont {V.~L.}\ \bibnamefont {Korenev}}, \bibinfo {author}
  {\bibfnamefont {T.~L.}\ \bibnamefont {Reinecke}}, \ and\ \bibinfo {author}
  {\bibfnamefont {D.}~\bibnamefont {Gammon}},\ }\href {\doibase
  10.1103/PhysRevLett.97.197202} {\bibfield  {journal} {\bibinfo  {journal}
  {Phys. Rev. Lett.}\ }\textbf {\bibinfo {volume} {97}},\ \bibinfo {pages}
  {197202} (\bibinfo {year} {2006})}\BibitemShut {NoStop}%
\bibitem [{\citenamefont {Climente}\ \emph {et~al.}(2008)\citenamefont
  {Climente}, \citenamefont {Korkusinski}, \citenamefont {Goldoni},\ and\
  \citenamefont {Hawrylak}}]{climente_theory_2008}%
  \BibitemOpen
  \bibfield  {author} {\bibinfo {author} {\bibfnamefont {J.~I.}\ \bibnamefont
  {Climente}}, \bibinfo {author} {\bibfnamefont {M.}~\bibnamefont
  {Korkusinski}}, \bibinfo {author} {\bibfnamefont {G.}~\bibnamefont
  {Goldoni}}, \ and\ \bibinfo {author} {\bibfnamefont {P.}~\bibnamefont
  {Hawrylak}},\ }\href {\doibase 10.1103/PhysRevB.78.115323} {\bibfield
  {journal} {\bibinfo  {journal} {Phys. Rev. B}\ }\textbf {\bibinfo {volume}
  {78}},\ \bibinfo {pages} {115323} (\bibinfo {year} {2008})}\BibitemShut
  {NoStop}%
\bibitem [{\citenamefont {Ma}\ \emph {et~al.}(2016)\citenamefont {Ma},
  \citenamefont {Bryant},\ and\ \citenamefont {Doty}}]{ma_hole_2016}%
  \BibitemOpen
  \bibfield  {author} {\bibinfo {author} {\bibfnamefont {X.}~\bibnamefont
  {Ma}}, \bibinfo {author} {\bibfnamefont {G.~W.}\ \bibnamefont {Bryant}}, \
  and\ \bibinfo {author} {\bibfnamefont {M.~F.}\ \bibnamefont {Doty}},\ }\href
  {\doibase 10.1103/PhysRevB.93.245402} {\bibfield  {journal} {\bibinfo
  {journal} {Phys. Rev. B}\ }\textbf {\bibinfo {volume} {93}},\ \bibinfo
  {pages} {245402} (\bibinfo {year} {2016})}\BibitemShut {NoStop}%
\bibitem [{\citenamefont {Janotti}\ \emph {et~al.}(2002)\citenamefont
  {Janotti}, \citenamefont {Wei},\ and\ \citenamefont
  {Zhang}}]{janotti_theoretical_2002}%
  \BibitemOpen
  \bibfield  {author} {\bibinfo {author} {\bibfnamefont {A.}~\bibnamefont
  {Janotti}}, \bibinfo {author} {\bibfnamefont {S.-H.}\ \bibnamefont {Wei}}, \
  and\ \bibinfo {author} {\bibfnamefont {S.~B.}\ \bibnamefont {Zhang}},\ }\href
  {\doibase 10.1103/PhysRevB.65.115203} {\bibfield  {journal} {\bibinfo
  {journal} {Phys. Rev. B}\ }\textbf {\bibinfo {volume} {65}},\ \bibinfo
  {pages} {115203} (\bibinfo {year} {2002})}\BibitemShut {NoStop}%
\bibitem [{\citenamefont {Yoshida}\ \emph {et~al.}(2003)\citenamefont
  {Yoshida}, \citenamefont {Kita}, \citenamefont {Wada},\ and\ \citenamefont
  {Oe}}]{yoshida_temperature_2003}%
  \BibitemOpen
  \bibfield  {author} {\bibinfo {author} {\bibfnamefont {J.}~\bibnamefont
  {Yoshida}}, \bibinfo {author} {\bibfnamefont {T.}~\bibnamefont {Kita}},
  \bibinfo {author} {\bibfnamefont {O.}~\bibnamefont {Wada}}, \ and\ \bibinfo
  {author} {\bibfnamefont {K.}~\bibnamefont {Oe}},\ }\href {\doibase
  10.1143/JJAP.42.371} {\bibfield  {journal} {\bibinfo  {journal} {Jpn. J.
  Appl. Phys.}\ }\textbf {\bibinfo {volume} {42}},\ \bibinfo {pages} {371}
  (\bibinfo {year} {2003})}\BibitemShut {NoStop}%
\bibitem [{\citenamefont {Zhang}\ \emph {et~al.}(2005)\citenamefont {Zhang},
  \citenamefont {Mascarenhas},\ and\ \citenamefont
  {Wang}}]{zhang_similar_2005}%
  \BibitemOpen
  \bibfield  {author} {\bibinfo {author} {\bibfnamefont {Y.}~\bibnamefont
  {Zhang}}, \bibinfo {author} {\bibfnamefont {A.}~\bibnamefont {Mascarenhas}},
  \ and\ \bibinfo {author} {\bibfnamefont {L.-W.}\ \bibnamefont {Wang}},\
  }\href {\doibase 10.1103/PhysRevB.71.155201} {\bibfield  {journal} {\bibinfo
  {journal} {Phys. Rev. B}\ }\textbf {\bibinfo {volume} {71}},\ \bibinfo
  {pages} {155201} (\bibinfo {year} {2005})}\BibitemShut {NoStop}%
\bibitem [{\citenamefont {Alberi}\ \emph {et~al.}(2007)\citenamefont {Alberi},
  \citenamefont {Wu}, \citenamefont {Walukiewicz}, \citenamefont {Yu},
  \citenamefont {Dubon}, \citenamefont {Watkins}, \citenamefont {Wang},
  \citenamefont {Liu}, \citenamefont {Cho},\ and\ \citenamefont
  {Furdyna}}]{alberi_valence-band_2007}%
  \BibitemOpen
  \bibfield  {author} {\bibinfo {author} {\bibfnamefont {K.}~\bibnamefont
  {Alberi}}, \bibinfo {author} {\bibfnamefont {J.}~\bibnamefont {Wu}}, \bibinfo
  {author} {\bibfnamefont {W.}~\bibnamefont {Walukiewicz}}, \bibinfo {author}
  {\bibfnamefont {K.~M.}\ \bibnamefont {Yu}}, \bibinfo {author} {\bibfnamefont
  {O.~D.}\ \bibnamefont {Dubon}}, \bibinfo {author} {\bibfnamefont {S.~P.}\
  \bibnamefont {Watkins}}, \bibinfo {author} {\bibfnamefont {C.~X.}\
  \bibnamefont {Wang}}, \bibinfo {author} {\bibfnamefont {X.}~\bibnamefont
  {Liu}}, \bibinfo {author} {\bibfnamefont {Y.-J.}\ \bibnamefont {Cho}}, \ and\
  \bibinfo {author} {\bibfnamefont {J.}~\bibnamefont {Furdyna}},\ }\href
  {\doibase 10.1103/PhysRevB.75.045203} {\bibfield  {journal} {\bibinfo
  {journal} {Phys. Rev. B}\ }\textbf {\bibinfo {volume} {75}},\ \bibinfo
  {pages} {045203} (\bibinfo {year} {2007})}\BibitemShut {NoStop}%
\bibitem [{\citenamefont {Deng}\ \emph {et~al.}(2010)\citenamefont {Deng},
  \citenamefont {Li}, \citenamefont {Li}, \citenamefont {Peng}, \citenamefont
  {Xia}, \citenamefont {Wang},\ and\ \citenamefont {Wei}}]{deng_band_2010}%
  \BibitemOpen
  \bibfield  {author} {\bibinfo {author} {\bibfnamefont {H.-X.}\ \bibnamefont
  {Deng}}, \bibinfo {author} {\bibfnamefont {J.}~\bibnamefont {Li}}, \bibinfo
  {author} {\bibfnamefont {S.-S.}\ \bibnamefont {Li}}, \bibinfo {author}
  {\bibfnamefont {H.}~\bibnamefont {Peng}}, \bibinfo {author} {\bibfnamefont
  {J.-B.}\ \bibnamefont {Xia}}, \bibinfo {author} {\bibfnamefont {L.-W.}\
  \bibnamefont {Wang}}, \ and\ \bibinfo {author} {\bibfnamefont {S.-H.}\
  \bibnamefont {Wei}},\ }\href {\doibase 10.1103/PhysRevB.82.193204} {\bibfield
   {journal} {\bibinfo  {journal} {Phys. Rev. B}\ }\textbf {\bibinfo {volume}
  {82}},\ \bibinfo {pages} {193204} (\bibinfo {year} {2010})}\BibitemShut
  {NoStop}%
\bibitem [{\citenamefont {Usman}\ \emph {et~al.}(2011)\citenamefont {Usman},
  \citenamefont {Broderick}, \citenamefont {Lindsay},\ and\ \citenamefont {O{
  extquoteright}Reilly}}]{usman_tight-binding_2011}%
  \BibitemOpen
  \bibfield  {author} {\bibinfo {author} {\bibfnamefont {M.}~\bibnamefont
  {Usman}}, \bibinfo {author} {\bibfnamefont {C.~A.}\ \bibnamefont
  {Broderick}}, \bibinfo {author} {\bibfnamefont {A.}~\bibnamefont {Lindsay}},
  \ and\ \bibinfo {author} {\bibfnamefont {E.~P.}\ \bibnamefont {O{
  extquoteright}Reilly}},\ }\href {\doibase 10.1103/PhysRevB.84.245202}
  {\bibfield  {journal} {\bibinfo  {journal} {Phys. Rev. B}\ }\textbf {\bibinfo
  {volume} {84}},\ \bibinfo {pages} {245202} (\bibinfo {year}
  {2011})}\BibitemShut {NoStop}%
\bibitem [{\citenamefont {Batool}\ \emph {et~al.}(2012)\citenamefont {Batool},
  \citenamefont {Hild}, \citenamefont {Hosea}, \citenamefont {Lu},
  \citenamefont {Tiedje},\ and\ \citenamefont
  {Sweeney}}]{batool_electronic_2012}%
  \BibitemOpen
  \bibfield  {author} {\bibinfo {author} {\bibfnamefont {Z.}~\bibnamefont
  {Batool}}, \bibinfo {author} {\bibfnamefont {K.}~\bibnamefont {Hild}},
  \bibinfo {author} {\bibfnamefont {T.~J.~C.}\ \bibnamefont {Hosea}}, \bibinfo
  {author} {\bibfnamefont {X.}~\bibnamefont {Lu}}, \bibinfo {author}
  {\bibfnamefont {T.}~\bibnamefont {Tiedje}}, \ and\ \bibinfo {author}
  {\bibfnamefont {S.~J.}\ \bibnamefont {Sweeney}},\ }\href {\doibase
  10.1063/1.4728028} {\bibfield  {journal} {\bibinfo  {journal} {Journal of
  Applied Physics}\ }\textbf {\bibinfo {volume} {111}},\ \bibinfo {pages}
  {113108} (\bibinfo {year} {2012})}\BibitemShut {NoStop}%
\bibitem [{\citenamefont {Usman}\ \emph {et~al.}(2013)\citenamefont {Usman},
  \citenamefont {Broderick}, \citenamefont {Batool}, \citenamefont {Hild},
  \citenamefont {Hosea}, \citenamefont {Sweeney},\ and\ \citenamefont {O{
  extquoteright}Reilly}}]{usman_impact_2013}%
  \BibitemOpen
  \bibfield  {author} {\bibinfo {author} {\bibfnamefont {M.}~\bibnamefont
  {Usman}}, \bibinfo {author} {\bibfnamefont {C.~A.}\ \bibnamefont
  {Broderick}}, \bibinfo {author} {\bibfnamefont {Z.}~\bibnamefont {Batool}},
  \bibinfo {author} {\bibfnamefont {K.}~\bibnamefont {Hild}}, \bibinfo {author}
  {\bibfnamefont {T.~J.~C.}\ \bibnamefont {Hosea}}, \bibinfo {author}
  {\bibfnamefont {S.~J.}\ \bibnamefont {Sweeney}}, \ and\ \bibinfo {author}
  {\bibfnamefont {E.~P.}\ \bibnamefont {O{ extquoteright}Reilly}},\ }\href
  {\doibase 10.1103/PhysRevB.87.115104} {\bibfield  {journal} {\bibinfo
  {journal} {Phys. Rev. B}\ }\textbf {\bibinfo {volume} {87}},\ \bibinfo
  {pages} {115104} (\bibinfo {year} {2013})}\BibitemShut {NoStop}%
\bibitem [{\citenamefont {Bannow}\ \emph {et~al.}(2016)\citenamefont {Bannow},
  \citenamefont {Rubel}, \citenamefont {Badescu}, \citenamefont {Rosenow},
  \citenamefont {Hader}, \citenamefont {Moloney}, \citenamefont {Tonner},\ and\
  \citenamefont {Koch}}]{bannow_configuration_2016}%
  \BibitemOpen
  \bibfield  {author} {\bibinfo {author} {\bibfnamefont {L.~C.}\ \bibnamefont
  {Bannow}}, \bibinfo {author} {\bibfnamefont {O.}~\bibnamefont {Rubel}},
  \bibinfo {author} {\bibfnamefont {S.~C.}\ \bibnamefont {Badescu}}, \bibinfo
  {author} {\bibfnamefont {P.}~\bibnamefont {Rosenow}}, \bibinfo {author}
  {\bibfnamefont {J.}~\bibnamefont {Hader}}, \bibinfo {author} {\bibfnamefont
  {J.~V.}\ \bibnamefont {Moloney}}, \bibinfo {author} {\bibfnamefont
  {R.}~\bibnamefont {Tonner}}, \ and\ \bibinfo {author} {\bibfnamefont {S.~W.}\
  \bibnamefont {Koch}},\ }\href {\doibase 10.1103/PhysRevB.93.205202}
  {\bibfield  {journal} {\bibinfo  {journal} {Phys. Rev. B}\ }\textbf {\bibinfo
  {volume} {93}},\ \bibinfo {pages} {205202} (\bibinfo {year}
  {2016})}\BibitemShut {NoStop}%
\bibitem [{\citenamefont {Bannow}\ \emph {et~al.}(2017)\citenamefont {Bannow},
  \citenamefont {Badescu}, \citenamefont {Hader}, \citenamefont {Moloney},\
  and\ \citenamefont {Koch}}]{bannow_valence_2017}%
  \BibitemOpen
  \bibfield  {author} {\bibinfo {author} {\bibfnamefont {L.~C.}\ \bibnamefont
  {Bannow}}, \bibinfo {author} {\bibfnamefont {S.~C.}\ \bibnamefont {Badescu}},
  \bibinfo {author} {\bibfnamefont {J.}~\bibnamefont {Hader}}, \bibinfo
  {author} {\bibfnamefont {J.~V.}\ \bibnamefont {Moloney}}, \ and\ \bibinfo
  {author} {\bibfnamefont {S.~W.}\ \bibnamefont {Koch}},\ }\href {\doibase
  10.1063/1.5005156} {\bibfield  {journal} {\bibinfo  {journal} {Appl. Phys.
  Lett.}\ }\textbf {\bibinfo {volume} {111}},\ \bibinfo {pages} {182103}
  (\bibinfo {year} {2017})}\BibitemShut {NoStop}%
\bibitem [{\citenamefont {Wasilewski}\ \emph {et~al.}(1999)\citenamefont
  {Wasilewski}, \citenamefont {Fafard},\ and\ \citenamefont
  {McCaffrey}}]{wasilewski_size_1999}%
  \BibitemOpen
  \bibfield  {author} {\bibinfo {author} {\bibfnamefont {Z.~R.}\ \bibnamefont
  {Wasilewski}}, \bibinfo {author} {\bibfnamefont {S.}~\bibnamefont {Fafard}},
  \ and\ \bibinfo {author} {\bibfnamefont {J.~P.}\ \bibnamefont {McCaffrey}},\
  }\href {\doibase 10.1016/S0022-0248(98)01539-5} {\bibfield  {journal}
  {\bibinfo  {journal} {Journal of Crystal Growth}\ }\textbf {\bibinfo {volume}
  {201-202}},\ \bibinfo {pages} {1131} (\bibinfo {year} {1999})}\BibitemShut
  {NoStop}%
\bibitem [{\citenamefont {Scheibner}\ \emph {et~al.}(2009)\citenamefont
  {Scheibner}, \citenamefont {Bracker}, \citenamefont {Kim},\ and\
  \citenamefont {Gammon}}]{scheibner_essential_2009}%
  \BibitemOpen
  \bibfield  {author} {\bibinfo {author} {\bibfnamefont {M.}~\bibnamefont
  {Scheibner}}, \bibinfo {author} {\bibfnamefont {A.~S.}\ \bibnamefont
  {Bracker}}, \bibinfo {author} {\bibfnamefont {D.}~\bibnamefont {Kim}}, \ and\
  \bibinfo {author} {\bibfnamefont {D.}~\bibnamefont {Gammon}},\ }\href
  {\doibase 10.1016/j.ssc.2009.04.039} {\bibfield  {journal} {\bibinfo
  {journal} {Solid State Communications}\ }\bibinfo {series} {Fundamental
  {Phenomena} and {Applications} of {Quantum} {Dots}},\ \textbf {\bibinfo
  {volume} {149}},\ \bibinfo {pages} {1427} (\bibinfo {year}
  {2009})}\BibitemShut {NoStop}%
\bibitem [{\citenamefont {Harrison}(1973)}]{harrison_bond-orbital_1973}%
  \BibitemOpen
  \bibfield  {author} {\bibinfo {author} {\bibfnamefont {W.~A.}\ \bibnamefont
  {Harrison}},\ }\href {\doibase 10.1103/PhysRevB.8.4487} {\bibfield  {journal}
  {\bibinfo  {journal} {Phys. Rev. B}\ }\textbf {\bibinfo {volume} {8}},\
  \bibinfo {pages} {4487} (\bibinfo {year} {1973})}\BibitemShut {NoStop}%
\bibitem [{\citenamefont {Harrison}\ and\ \citenamefont
  {Ciraci}(1974)}]{harrison_bond-orbital_1974}%
  \BibitemOpen
  \bibfield  {author} {\bibinfo {author} {\bibfnamefont {W.~A.}\ \bibnamefont
  {Harrison}}\ and\ \bibinfo {author} {\bibfnamefont {S.}~\bibnamefont
  {Ciraci}},\ }\href {\doibase 10.1103/PhysRevB.10.1516} {\bibfield  {journal}
  {\bibinfo  {journal} {Phys. Rev. B}\ }\textbf {\bibinfo {volume} {10}},\
  \bibinfo {pages} {1516} (\bibinfo {year} {1974})}\BibitemShut {NoStop}%
\bibitem [{\citenamefont {Pantelides}\ and\ \citenamefont
  {Harrison}(1975)}]{pantelides_structure_1975}%
  \BibitemOpen
  \bibfield  {author} {\bibinfo {author} {\bibfnamefont {S.~T.}\ \bibnamefont
  {Pantelides}}\ and\ \bibinfo {author} {\bibfnamefont {W.~A.}\ \bibnamefont
  {Harrison}},\ }\href {\doibase 10.1103/PhysRevB.11.3006} {\bibfield
  {journal} {\bibinfo  {journal} {Phys. Rev. B}\ }\textbf {\bibinfo {volume}
  {11}},\ \bibinfo {pages} {3006} (\bibinfo {year} {1975})}\BibitemShut
  {NoStop}%
\bibitem [{\citenamefont {Vogl}\ \emph {et~al.}(1983)\citenamefont {Vogl},
  \citenamefont {Hjalmarson},\ and\ \citenamefont
  {Dow}}]{vogl_semi-empirical_1983}%
  \BibitemOpen
  \bibfield  {author} {\bibinfo {author} {\bibfnamefont {P.}~\bibnamefont
  {Vogl}}, \bibinfo {author} {\bibfnamefont {H.~P.}\ \bibnamefont
  {Hjalmarson}}, \ and\ \bibinfo {author} {\bibfnamefont {J.~D.}\ \bibnamefont
  {Dow}},\ }\href {\doibase 10.1016/0022-3697(83)90064-1} {\bibfield  {journal}
  {\bibinfo  {journal} {Journal of Physics and Chemistry of Solids}\ }\textbf
  {\bibinfo {volume} {44}},\ \bibinfo {pages} {365} (\bibinfo {year}
  {1983})}\BibitemShut {NoStop}%
\bibitem [{\citenamefont {Graf}\ and\ \citenamefont
  {Vogl}(1995)}]{graf_electromagnetic_1995}%
  \BibitemOpen
  \bibfield  {author} {\bibinfo {author} {\bibfnamefont {M.}~\bibnamefont
  {Graf}}\ and\ \bibinfo {author} {\bibfnamefont {P.}~\bibnamefont {Vogl}},\
  }\href {\doibase 10.1103/PhysRevB.51.4940} {\bibfield  {journal} {\bibinfo
  {journal} {Phys. Rev. B}\ }\textbf {\bibinfo {volume} {51}},\ \bibinfo
  {pages} {4940} (\bibinfo {year} {1995})}\BibitemShut {NoStop}%
\bibitem [{\citenamefont {Pryor}\ \emph {et~al.}(1998)\citenamefont {Pryor},
  \citenamefont {Kim}, \citenamefont {Wang}, \citenamefont {Williamson},\ and\
  \citenamefont {Zunger}}]{pryor_comparison_1998}%
  \BibitemOpen
  \bibfield  {author} {\bibinfo {author} {\bibfnamefont {C.}~\bibnamefont
  {Pryor}}, \bibinfo {author} {\bibfnamefont {J.}~\bibnamefont {Kim}}, \bibinfo
  {author} {\bibfnamefont {L.~W.}\ \bibnamefont {Wang}}, \bibinfo {author}
  {\bibfnamefont {A.~J.}\ \bibnamefont {Williamson}}, \ and\ \bibinfo {author}
  {\bibfnamefont {A.}~\bibnamefont {Zunger}},\ }\href {\doibase
  10.1063/1.366631} {\bibfield  {journal} {\bibinfo  {journal} {Journal of
  Applied Physics}\ }\textbf {\bibinfo {volume} {83}},\ \bibinfo {pages} {2548}
  (\bibinfo {year} {1998})}\BibitemShut {NoStop}%
\bibitem [{\citenamefont {Pashartis}\ and\ \citenamefont
  {Rubel}(2017)}]{pashartis_localization_2017}%
  \BibitemOpen
  \bibfield  {author} {\bibinfo {author} {\bibfnamefont {C.}~\bibnamefont
  {Pashartis}}\ and\ \bibinfo {author} {\bibfnamefont {O.}~\bibnamefont
  {Rubel}},\ }\href {\doibase 10.1103/PhysRevApplied.7.064011} {\bibfield
  {journal} {\bibinfo  {journal} {Phys. Rev. Applied}\ }\textbf {\bibinfo
  {volume} {7}},\ \bibinfo {pages} {064011} (\bibinfo {year}
  {2017})}\BibitemShut {NoStop}%
\bibitem [{\citenamefont {Lindsay}\ and\ \citenamefont
  {O'Reilly}(1999)}]{lindsay_theory_1999}%
  \BibitemOpen
  \bibfield  {author} {\bibinfo {author} {\bibfnamefont {A.}~\bibnamefont
  {Lindsay}}\ and\ \bibinfo {author} {\bibfnamefont {E.~P.}\ \bibnamefont
  {O'Reilly}},\ }\href {\doibase 10.1016/S0038-1098(99)00361-0} {\bibfield
  {journal} {\bibinfo  {journal} {Solid State Communications}\ }\textbf
  {\bibinfo {volume} {112}},\ \bibinfo {pages} {443} (\bibinfo {year}
  {1999})}\BibitemShut {NoStop}%
\bibitem [{\citenamefont {Laukkanen}\ \emph {et~al.}(2017)\citenamefont
  {Laukkanen}, \citenamefont {Punkkinen}, \citenamefont {Lahti}, \citenamefont
  {Puustinen}, \citenamefont {Tuominen}, \citenamefont {Hilska}, \citenamefont
  {M{\"a}kel{\"a}}, \citenamefont {Dahl}, \citenamefont {Yasir}, \citenamefont
  {Kuzmin}, \citenamefont {Osiecki}, \citenamefont {Schulte}, \citenamefont
  {Guina},\ and\ \citenamefont {Kokko}}]{laukkanen_local_2017}%
  \BibitemOpen
  \bibfield  {author} {\bibinfo {author} {\bibfnamefont {P.}~\bibnamefont
  {Laukkanen}}, \bibinfo {author} {\bibfnamefont {M.~P.~J.}\ \bibnamefont
  {Punkkinen}}, \bibinfo {author} {\bibfnamefont {A.}~\bibnamefont {Lahti}},
  \bibinfo {author} {\bibfnamefont {J.}~\bibnamefont {Puustinen}}, \bibinfo
  {author} {\bibfnamefont {M.}~\bibnamefont {Tuominen}}, \bibinfo {author}
  {\bibfnamefont {J.}~\bibnamefont {Hilska}}, \bibinfo {author} {\bibfnamefont
  {J.}~\bibnamefont {M{\"a}kel{\"a}}}, \bibinfo {author} {\bibfnamefont
  {J.}~\bibnamefont {Dahl}}, \bibinfo {author} {\bibfnamefont {M.}~\bibnamefont
  {Yasir}}, \bibinfo {author} {\bibfnamefont {M.}~\bibnamefont {Kuzmin}},
  \bibinfo {author} {\bibfnamefont {J.~R.}\ \bibnamefont {Osiecki}}, \bibinfo
  {author} {\bibfnamefont {K.}~\bibnamefont {Schulte}}, \bibinfo {author}
  {\bibfnamefont {M.}~\bibnamefont {Guina}}, \ and\ \bibinfo {author}
  {\bibfnamefont {K.}~\bibnamefont {Kokko}},\ }\href {\doibase
  10.1016/j.apsusc.2016.11.009} {\bibfield  {journal} {\bibinfo  {journal}
  {Applied Surface Science}\ }\textbf {\bibinfo {volume} {396}},\ \bibinfo
  {pages} {688} (\bibinfo {year} {2017})}\BibitemShut {NoStop}%
\bibitem [{\citenamefont {Fluegel}\ \emph {et~al.}(2006)\citenamefont
  {Fluegel}, \citenamefont {Francoeur}, \citenamefont {Mascarenhas},
  \citenamefont {Tixier}, \citenamefont {Young},\ and\ \citenamefont
  {Tiedje}}]{fluegel_giant_2006}%
  \BibitemOpen
  \bibfield  {author} {\bibinfo {author} {\bibfnamefont {B.}~\bibnamefont
  {Fluegel}}, \bibinfo {author} {\bibfnamefont {S.}~\bibnamefont {Francoeur}},
  \bibinfo {author} {\bibfnamefont {A.}~\bibnamefont {Mascarenhas}}, \bibinfo
  {author} {\bibfnamefont {S.}~\bibnamefont {Tixier}}, \bibinfo {author}
  {\bibfnamefont {E.~C.}\ \bibnamefont {Young}}, \ and\ \bibinfo {author}
  {\bibfnamefont {T.}~\bibnamefont {Tiedje}},\ }\href {\doibase
  10.1103/PhysRevLett.97.067205} {\bibfield  {journal} {\bibinfo  {journal}
  {Phys. Rev. Lett.}\ }\textbf {\bibinfo {volume} {97}},\ \bibinfo {pages}
  {067205} (\bibinfo {year} {2006})}\BibitemShut {NoStop}%
\end{thebibliography}%

\end{document}